\documentclass[]{aa}

\usepackage{graphicx}
\usepackage{txfonts}
\usepackage{subfig}
\usepackage{textgreek}
\usepackage{multirow}
\usepackage{mathtools}
\usepackage{color}
\usepackage{lmodern}
\usepackage{epstopdf}
\usepackage{xcolor}
\usepackage[draft]{hyperref}

\bibpunct{(}{)}{;}{a}{}{,}

\newcommand{\FIG}[1]{}

\def\mso{\,{\rm M}_\odot}

\def\lso{\,{\rm L}_\odot}

\def\kms{\, {\rm km}\, {\rm s}^{-1}}

\def\msoy{\, \mso~{\rm yr}^{-1}}

\begin{document}


   \title{A detailed view on the circumstellar environment of the M-type AGB star EP Aquarii}

   \subtitle{I. High-resolution ALMA and SPHERE observations}

   \author{Ward Homan
          \inst{1}
          \and
          Emily Cannon
          \inst{1}
          \and
          Miguel Montarg\`{e}s
          \inst{1}
          \and
          Anita M. S. Richards
          \inst{2}
          \and
          Tom J. Millar
          \inst{3}
          \and
          Leen Decin
          \inst{1,4}
          }

   \offprints{W. Homan}          
          
   \institute{$^{\rm 1}\ $Institute of Astronomy, KU Leuven, Celestijnenlaan 200D B2401, 3001 Leuven, Belgium \\
             $^{\rm 2}\ $JBCA, Department Physics and Astronomy, University of Manchester, Manchester M13 9PL, UK \\
             $^{\rm 3}\ $Astrophysics Research Centre, School of Mathematics and Physics, Queen’s University Belfast, Belfast BT7 1NN, UK \\
             $^{\rm 4}\ $School of Chemistry, University of Leeds, Leeds LS2 9JT, UK \\
             }

   \date{Received <date> / Accepted <date>}
 
   \abstract  
   {Cool evolved stars are known to be significant contributors to the enrichment of the interstellar medium through their dense and dusty stellar winds. High resolution observations of these outflows have shown them to possess high degrees of morphological complexity. We observed the asymptotic giant branch (AGB) star EP Aquarii with ALMA in band 6 and VLT/SPHERE/ZIMPOL in four filters the visible. Both instruments had an angular resolution of 0.025''. These are follow-up observations to the lower-resolution 2016 ALMA analysis of EP Aquarii, which revealed that its wind possesses a nearly face-on, spiral-harbouring equatorial density enhancement, with a nearly pole-on bi-conical outflow. At the base of the spiral, the SiO emission revealed a distinct emission void approximately 0.4'' to the west of the continuum brightness peak, which was proposed to be linked to the presence of a companion. The new ALMA data better resolve the inner wind and reveal that its morphology as observed in CO is consistent with hydrodynamical companion-induced perturbations. Assuming that photodissociation by the UV-field of the companion is responsible for the emission void in SiO, we deduced the spectral properties of the tentative companion from the size of the hole. We conclude that the most probable companion candidate is a white dwarf with a mass between 0.65 and 0.8$\mso$, though a solar-like companion could not be definitively excluded. The radial SiO emission shows periodic, low-amplitude perturbations. We tentatively propose that they could be the consequence of the interaction of the AGB wind with another much closer low-mass companion. The polarised SPHERE/ZIMPOL data show a circular signal surrounding the AGB star with a radius of $\sim$0.1''. Decreased signal along a PA of 138$^\circ$ suggests that the dust is confined to an inclined ring-like structure, consistent with the previously determined wind morphology.}
   

   \maketitle



\section{Introduction}

Near the end of their lives, all stars in the Universe initially more massive than 0.8 $\mso$ lose 40 $-$ 80\% of their mass via a stellar wind. For stars less than 8 $\mso$, this phase of intense mass-loss is known as the asymptotic giant Branch (AGB) phase. The mass-loss process close to the star is complex \citep[e.g.][]{Bladh2015} but, beyond a few stellar radii, the radial force on an AGB wind is thought to be due to radiation pressure on solid-state dust grains formed in the cool and extended atmospheres of the AGB star\citep{Hofner2018}. It has typical mass-loss rate values between $10^{-8} - 10^{-4} \msoy$, and velocities between 5 $-$ 20 $\kms$ \citep{Ramstedt2008}. For mass-loss rates above $\sim10^{-7}\msoy$, the associated time scale for stars to shed their envelope is much shorter than the nuclear burning time scale, implying that mass loss dominates and/or determines all further evolution \citep{McDonald2018}.

Historically, these winds were believed to possess a uniformly spherical geometry \citep{Neri1998}. Hence, they were also modelled as such, and most stellar wind properties presented in the literature (such as wind mass-loss rates) were derived under this assumption. However, recent observations have revealed that in fact most, if not all, AGB nebulae harbour a high degree of morphological complexity. Among these structural curiousities are, for example, bipolar structures, arcs, shells, clumps, spirals, tori, bubbles, and rotating discs \citep{Decin2020,Bowers1990,Cox2012,Maercker2012,Decin2012,Balick2013,Ramstedt2014,Kervella2015,Decin2015,Kervella2016}. A range of dynamical processes that operate on different length and time scales have been proposed to explain some of these structures \citep{Woitke2006b, Cox2012,VanMarle2014,Chen2016}. However, the frequent occurrence of cylindrical morphologies among planetary nebulae (PNe), their presence among the post-AGB PN predecessors \citep{Bujarrabal2015}, and the recent first detections of comparable morphologies around AGB stars \citep{Kervella2016,Doan2017,Homan2018} are steadily shifting the scientific consensus towards binary activity as the dominant shaping mechanism.

Most high-resolution observations of the circumstellar environments of AGB stars exhibit strong indications that binarity is the source of the spectrum of perturbations of the circumstellar gas and dust distributions. Furthermore, it has been shown that the misinterpretation of companion-perturbed AGB winds as smooth nebulae affects derived mass-loss rates \citep{Homan2015,Decin2019}. Hence, it is important to perform detailed observations, analyses, and modelling of the targets that exhibit deviations from spherical symmetry in order to obtain a better grasp on the physical and chemical conditions that dominate the wind-launching and wind-companion interaction zones.

Mass is one of the decisive evolutionary properties of stars in general. The rate of mass transport away from the stellar mantle by these stellar winds, which varies as a function of age, governs the change of the stellar mass with time, which in turn has decisive repercussions on the evolutionary path of the star and nucleosynthesis in its core. Hence, a proper grasp on stellar, and by extension galactic chemical, evolution cannot be achieved without a detailed understanding of the wind physics over a star's life cycle.

\begin{table*}
        \caption{The main line in each spw of the 2019 ALMA TM0 data.}
        \centering          
        \label{spw}
        \begin{tabular}{clllccr}
        \hline\hline
        \noalign{\smallskip}
spw & Objective & Transition & Line rest freq. & Chan. velocity & Noise rms & Beam size, PA\\
& & & (GHz) & width (km\,s$ ^{-1}$) & (mJy) & (mas $\times$ mas, deg)\\
        \noalign{\smallskip}
        \hline    
        \noalign{\smallskip}
0 & $^{12}$CO & $J$=2$-$1 & 230.55 & 0.32 & 1.3 & $26\times24$, +35 \\
        \hline    
        \noalign{\smallskip}
1 & SO$_{\rm 2}$ & $J$=4$_{\rm 2,2}-$3$_{\rm 1,3}$ & 235.15 & 0.31 & 1.6 & $26\times23$, +38 \\
        \hline    
        \noalign{\smallskip}
2 & SiO & $J$=5$-$4 & 217.12 & 0.34 & 1.5 & $25\times23$, +20 \\
        \hline    
        \noalign{\smallskip}
3 & $^{13}$CO & $J$=2$-$1 & 220.399 & 0.33 & 1.2 & $24\times22$, +39 \\
        \hline
      \end{tabular}
\end{table*}

\begin{table*}
        \caption{Summary of parameters of continuum measurements. Int and Peak stand for the total flux density and the peak of a fitted two-dimensional Gaussian component.}
        \centering          
        \label{epoch}
        \begin{tabular}{lllllcccc}
        \hline\hline
        \noalign{\smallskip}
 Epoch & Config & Peak & Int & rms & Peak & Peak & Beam size & Beam PA\\
 & & (mJy) & (mJy) & (mJy) & Position (RA) & position (DEC) & (mas$\times$mas) & (deg) \\
        \noalign{\smallskip}
        \hline    
        \noalign{\smallskip}
2016-10-08 & ACA (7m) & 19.5 & 19.7  & 0.5  & 21:46:31.877  & -02:12:45.57  & 9340$\times$3710 & -80 \\
        \hline    
        \noalign{\smallskip}
2016-12-13 & TM2 (12m) & 17.58 & 18.64 & 0.05 & 21:46:31.880  & -02:12:45.62  & 880$\times$660   & -77 \\
        \hline    
        \noalign{\smallskip}
2016-10-08 & TM1 (12m) & 16.59 & 17.21 & 0.03 & 21:46:31.879  & -02:12:45.59  & 160$\times$150   & -24 \\
        \hline    
        \noalign{\smallskip}
2016       & All  & 17.89 & 17.89 & 0.03 & 21:46:31.8793 & -02.12.45.595 & 170$\times$150   & -34 \\
        \hline    
        \noalign{\smallskip}
2019-06-05 & TM0 (12m) & 16.10 & 21.35 & 0.03 & 21:46:31.8832 & -02:12:45.529 & 23$\times$21     & -22 \\
        \hline    
        \noalign{\smallskip}
2019       & All  & 16.32 & 20.95 & 0.02 & 21:46:31.8778 & -02:12:45.586 & 23$\times$22     & -15\\
        \hline
      \end{tabular}
\end{table*}

\begin{table*}
        \caption{Summary of the properties of the combined datacubes.}
        \centering          
        \label{combined}
        \begin{tabular}{llllcr}
        \hline\hline
        \noalign{\smallskip}
Species & Transition & Line rest freq. & Noise rms & Beam size, PA\\
& & (GHz) &  (mJy) & (mas $\times$ mas, deg)\\
        \noalign{\smallskip}
        \hline    
        \noalign{\smallskip}
$^{12}$CO & $J$=2$-$1 & 230.55  & 0.6 & $87\times79$, +15 \\
        \hline    
        \noalign{\smallskip}
SiO & $J$=5$-$4 & 217.12 & 0.7 & $85\times79$, +7 \\
        \hline    
      \end{tabular}
\end{table*}

It is in this context that we present the latest results on the circumstellar environment of the M-type AGB star EP Aquarii. For an overview of the general CSE properties we refer to \citet{Homan2018b,Hoai2019} and \citet{TuanAnh2019}, who discuss the CSE gas properties based on ALMA cycle 4 observations of the CSE at a resolution of $\sim$0.2''. We provide a short summary of the main results: EP Aquarii is an M-type, semi-regular variable star (spectral type M8III), and is currently in the AGB evolutionary phase. It has a pulsation period of 55 days \citep{Lebzelter2002}, and Hipparcos measured its distance to the Earth to be approximately 135 parsec \citep{Lebzelter2002}. It possesses an effective temperature of $T_*\simeq$ 3200K, a mass of $M_*\simeq1.7\,\mso$, and a luminosity is $\sim$ 4800$\,\lso$ \citep{Dumm1998}. It is known to have a radial velocity relative to the local standard of rest (lsr) of $v_{*} = -34\,\kms$, a mass-loss rate of $\sim$1.2$\times$10$^{\rm -7}\,\msoy$, and a terminal wind velocity of $\sim$11$\,\kms$ \citep{Nhung2015}. High spatial resolution observations in the last few years have revealed the circumstellar environment of EP Aqr to be extremely complex. ALMA observations of CO emission reveal that the wind of EP Aqr possesses a pronounced face-on equatorial density enhancement (EDE) that harbours a spiral. This EDE is encased in two, nearly pole-on outflow hemispheres that could be considered bi-conical, and that exhibit a multitude of nearly circular filamentary structures. In the inner wind, the CO emission patterns resemble the wind-Roche-lobe-overflow hydrodynamical models \citep{Mohamed2012}. The SiO emission within the same dataset reveals a peculiar void of emission about 0.45'' to the west of the continuum brightness peak (CBP) position, located at the base of the spiral, and coincident with the position of the companion in the obovementioned hydrodynamical models. As a consequence, the void has been suggested to be a local environment generated by a companion. Finally, the SiO emisison also exhibited enhanced gas dynamics that has been interpreted as high-velocity jets.

In the current work we adopt a distance of 135 pc \citep{Lebzelter2002} in order to retain consistency with the work by \citet{Homan2018b}. However, the literature also presents an alternatitve distance measurement made by GAIA of 124 pc \citep{Gaia2018}. All analyses presented in this work scale with the distance. Hence, if the GAIA distance is more accurate, this difference implies a reduction of all quantitative analyses by about 8\%.

In this first paper, we present the latest cycle 6 band 6 ALMA data of the wind of EP Aqr at a resolution that is finer by a factor ten, that has been combined with the older dataset. In addition, we also present VLT/SPHERE data on the target, showing the distribution of dust in the inner wind. Subsequent papers will present the three-dimensional hydrodynamical modelling of these data, and the three-dimensional radiative post-processing of the model. The paper is organised as follows: In Sect. \ref{ALMAobs} we give an overview of the ALMA observations and how they have been prepared for analysis, and continue with a similar report on the SPHERE/ZIMPOL data in Sect. \ref{SPHEREobs}. Subsequently, we discuss the nature of the continuum emission in Sect. \ref{continuum}. We proceed with an elaborate description of the ALMA molecular emission in Sect. \ref{molec}. These observational results are discussed in Sect. \ref{discus}. Finally, we summarise our findings in Sect. \ref{summ}.


\section{ALMA data acquisition and reduction} \label{ALMAobs}

\subsection{Observations}

EP Aquarii was observed by ALMA in Band 6 for project code 2018.1.00750.S (PI W. Homan) using the extended configuration C43-10, hereafter referred to as TM0. These observations complement data taken in three configurations at lower resolutions, described by \citet{Homan2018b}, and aspects of data reduction common to all observations are described there. EP Aqr was observed at R.A. 21:46:31.88046, Dec. --02:12:45.5471. The antenna configuration consisted of 46 antennae, and covered baseline lengths between 83 and 16196 m, which are sensitive to angular scales up to 0.48''. As the phase reference source we used J2156-0037 (position R.A. 21:56:14.758, Dec. 00:37:04.594), and the QSO targets J2232+1143 and J2253+1608 were used as bandpass calibrators. The target:phase reference switching cycle was about 80 sec, with about 50 sec on target, for a total of 99 minutes on target, and the angular separation was 3$^{\circ}$. J2206-0031 was observed as a check source at 22:06:43.282598 --00.31.02.49604, 246.5$^\circ$ from the reference source. During observations, the precipitable water vapour (PWV) was between 1 and 2.6 mm.

Listed in Table \ref{spw} are the four spectral windows (spw) that were observed, each with central frequencies at 217.128, 220.422, 230.562 and 235.176 GHz. Each spw has a width of 0.468692 GHz, divided into 1920 channels, resulting in a velocity channel width of $\sim0.32$ km s$^{-1}$, depending on frequency. 

\subsection{Data reduction and imaging}

The data were calibrated using the ALMA pipeline. We split out the pipeline-calibrated target data, adjusted to the Local Standard of Rest ($v_{\mathrm{LSR}}$) in the direction of the target, with one set of data at full spectral resolution and one averaged to 32 channels per spw, for continuum. Scanning the visibility spectra, we selected the line-free channels and used these to image the continuum, which were combined over all epochs. We then used this image as a starting model for self-calibration of phase and then amplitude. The corrections from the continuum self-calibration were applied also to the line data. The best TM0 continuum image has a noise rms level of 0.032 mJy, a 21 $\times$ 23 mas$^{\rm 2}$ beam size at a PA (PA) of --212$^{\circ}$. 

After continuum-subtraction, we imaged 100-150 channels of each spw, corresponding to the maximum velocity extent of the previously detected molecular emission, and used a pixel size of 0''.003 and a total image size of 6''. The rms noise, given in Table \ref{spw}, was measured in the channels devoid of signal. 

\subsection{Combination of multi-array data}

The combination of the lower-resolution data is described in \citet{Homan2018b}. We followed the same procedure to add the extended-configuration data to the combined dataset. At each epoch, the pointing position was adjusted for the proper motion of EP Aqr, based on the Hipparcos data available at proposal submission,(24.98$\pm$0.75, 19.54$\pm$0.33) mas yr$^{-1}$ \citep{VanLeeuwen2007}.

The positions of the continuum peaks at each epoch are given in Table \ref{epoch}. The extended configuration peak is offset by (40, 18) mas from the pointing position. Position errors due to noise are $<1$ mas and the check source's apparent position was displaced by $\sim$4 mas, mainly due to errors in phase transfer from the reference source. This suggests that the offset of (63, 9) mas noted by \citet{Homan2018b} was not entirely due to noise (although the uncertainty was greater at the lower resolution), but there was a small error in the input positions and/or proper motions. The recently published Gaia DR2 \citep{Gaia2018}, gives an epoch J2000 position of 21:46:31.84675 --02:12:45.9017, with an uncertainty of 0.46 and 0.39 mas, respectively, and with proper motions of 28.204$\pm$0.763, 18.794$\pm$0.682 mas yr$^{-1}$ in right ascension and declination, respectively. This predicts a position at epoch 2019.5 of 21:46:31.8834 ($\pm$0".015) and --02:12:45.535 ($\pm$0".0133). Our 2019 position is 21:46:31.8832 --02:12:45.529, an offset of (0.003'', 0.006''), which is within the Gaia uncertainties.

The relative error in the pointing positions between 2016 and 2019 due to the difference between Hipparcos and Gaia proper motions is $<$10 mas, which is less than the position uncertainties of all but the most recent data. Therefore, we did not adjust the positions, but combined the data aligned with the phase centre of the earliest observation. This proved sufficiently accurate to avoid imaging artefacts, but the 2019 TM0 position given above should be used for astrometry of the stellar position.

The continuum peak flux densities for all epochs are very similar at resolutions $<$few 100 mas, with a value of $\sim16.6\pm0.5$ mJy/beam, and a maximum change of 6\%. This is consistent with the ALMA flux scale accuracy of 7\%. The larger-scale continuum emission varies up to 20\% (see Table \ref{epoch}). These variations are probably intrinsic to the immediate stellar environment, and not calibration errors (see Sect. \ref{almacont}). Stellar variability will affect the immediate vicinity and radiatively excited lines, especially near the star, while for lines formed outside a few stellar radii the effects will be diluted and probably differ in different directions in an inhomogenous envelope. We therefore did not attempt to modify the line flux density scales.

We combined the line-free continuum from all epochs, with equal weight for each dataset. The parameters of the combined image are given in Table \ref{epoch}. We combined the full spectral resolution data for the two spectral windows with very similar coverage, around 217 and 230 GHz. The TM0 data were given a weight of 0.2 relative to the combined 2016 data, in order to give sufficient weight to the shorter spacings to detect extended emission (with imageable on scales up to 15'') and avoid deconvolution artefacts. With a pixel size of 5 mas, we imaged the channels around the SiO and CO lines using multi-scale clean, and the primary beam correction was applied. The rms noise, given in Table \ref{combined}, was measured in quiet channels.

\begin{table*}
\caption{SPHERE Observation log}             
\label{sphereobs}      
\centering          
\begin{tabular}{l l l c c c c c c c l }     
\hline       
\hline      
\noalign{\smallskip}                      
Star & UT 2018-08-17 & Filter & $\lambda$ [nm] & $\Delta \lambda$ & $\theta$ ["] & DIT[s] $\times$ NDIT & AM  & $\theta_\text{PSF}$ [mas] & Strehl & note \\ 
\hline 
\noalign{\smallskip}
EP Aqr   & 04:31:52  & CntH$\alpha$  & 644.9  & 4.1   & 0.60   & 5.0 $\times$ 12   & 1.085  & -  & 0.67 & -     \\
         &           & BH$\alpha$    & 655.6  & 5.5   & 0.60   & 5.0 $\times$ 12   & 1.085  & -  & 0.68 & -     \\
         & 05:17:31  & V             & 554    & 80.6  & 0.62   & 5.0 $\times$ 12   & 1.092  & -  & 0.55 & sat.  \\
         &           & Cnt748        & 747.4  & 20.6  & 0.62   & 5.0 $\times$ 12   & 1.092  & -  & 0.72 & sat.  \\
HD 1921  & 04:58:23  & CntH$\alpha$  & 644.9  & 4.1   & 0.71   & 5.0 $\times$ 8    & 1.208  & 27 & 0.24 & -     \\
         &           & BH$\alpha$    & 655.6  & 5.5   & 0.71   & 5.0 $\times$ 8    & 1.208  & 27 & 0.25 & -     \\
         & 05:55:26  & V             & 554    & 80.6  & 0.54   & 5.0 $\times$ 8    & 1.073  & 26 & 0.20 & ND\_1 \\
         &           & Cnt748        & 747.4  & 20.6  & 0.54   & 5.0 $\times$ 8    & 1.073  & 27 & 0.41 & ND\_1 \\      
\hline                 
\end{tabular}
\label{obstab}
\vspace{1ex}\\
\begin{flushleft}
\textbf{Note}: UT stands for universal time, $\theta$ is the visible seeing, AM gives the air mass, and $\theta_\text{PSF}$ is the FWHM of the PSF images. DIT gives the integration time of each frame and NDIT is the number of integrations. ND\_1 indicates the neutral density filter used. Sat. indicates CCD saturation. The Strehl ratios were obtained from the SPARTA (Standard Platform for Adaptive optics Real Time Applications) data and converted to the filter wavelength using the Marechal approximation.
\end{flushleft}
\end{table*} 

\begin{figure}[]
        \centering
        \includegraphics[width=8.5cm]{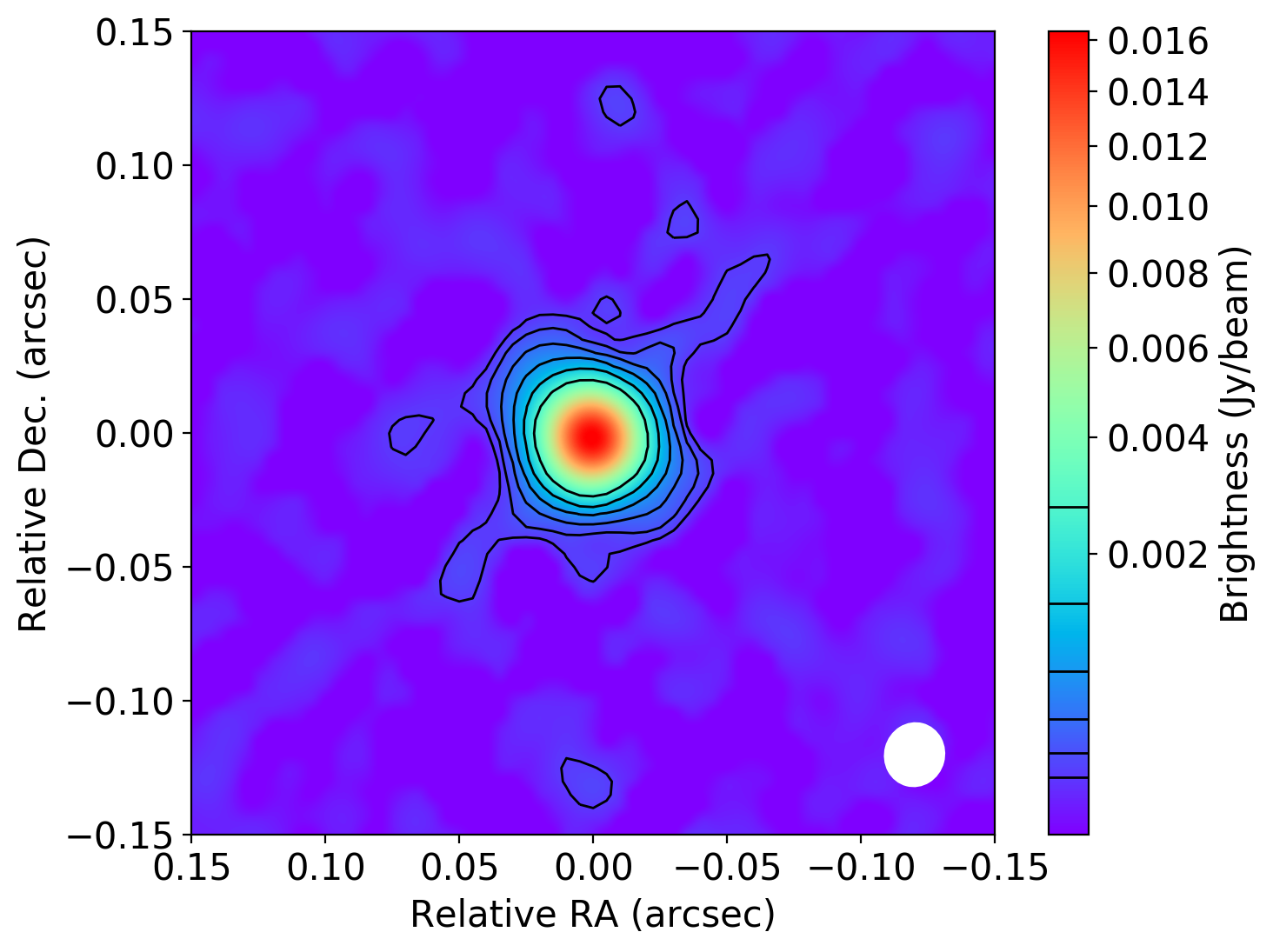}
        \caption{Continuum emission of the combined ALMA data on the EP Aqr system. Contours are drawn at 3, 6, 12, 24, 48 and 96 times the continuum rms noise value ($\sigma_{rms}$ = 2.83 $\times {\rm 10}^{\rm -5}$ Jy/beam). The offset position of the target with respect to the telescope pointing (ICRS 21:46:31.8475, -02:12:45.930) is due to transverse motion over the field of view. The ALMA beam is shown in the bottom right corner.
        \label{cont}} 
\end{figure}

\section{SPHERE data acquisition and reduction} \label{SPHEREobs}

We observed EP Aqr using SPHERE/ZIMPOL \citep{Beuzit2019,Schmid2018} in a Service Mode observing run (0102.D-0006(A)) in four filters, V, CntH$\alpha$, BH$\alpha$ and Cnt748, on August 17th 2018, along with HD1921 which was used as a point-spread function (PSF) calibrator star. 
The raw data was reduced using the ESO reflex data pipeline v0.24.0 \citep{Freudling2013}. 
The total observed intensity (I), polarised flux ($\sqrt{Q^2+U^2}$) and degree of linear polarisation ($\sqrt{Q^2+U^2} / I$) were computed from the Stokes parameter maps, with Q and U the Stokes parameters. 
The reduction products are shown in Fig. \ref{sphere}, which focuses on the innermost 0.3''$\times$ 0.3'' of the total field of view. 
At larger length-scales, the images are dominated by noise, as shown in Fig. \ref{sphererms} in the appendix. 
The observations using the V and Cnt748 filters are excluded from the analysis as the CCD is saturated for the science target. 
The PSF calibrator star is shown in Fig. \ref{calib} in the appendix.


\section{Continuum data description} \label{continuum}

In this section we present the continuum data obtained by the ALMA continuum subtraction procedure, and the SPHERE filters that show no saturation of the CCD.

\subsection{ALMA continuum emission} \label{almacont}

In Fig. \ref{cont} the ALMA continuum emission of the latest data combination is shown. The continuum peak is 16.32 mJy/beam, with a total flux within the 3$\sigma_{rms}$ ($\sigma_{rms}$ = 2.83$\times$10$^{\rm -5}$ Jy/beam) contour of 23.5 mJy, with a radius of approximately 40 milliarcseconds (mas). The emission is predominantly centrally condensed, with a slight north-east to south-west elongation, and with 2 symmetric filaments around the 135$^\circ$ position angle (PA, measured north-to-east) axis appearing in the faintest emission. These are sidelobe remnants of the combination, probably present due to the variability of the target between epochs of observation, which would limit the continuum dynamic range in the combined image. The same features can be recognised in the dirty beam shown in Fig. \ref{dirty} in the appendix. The resolution and signal-to-noise are good enough to deconvolve with tha ALMA beam, permitting the analysis of the geometry of the central source. Fitting it with a 2-D Gaussian results in a measured size of 14.0 $\pm$ 0.6 mas $\times$ 11.5 $\pm$ 0.7 mas, with a PA of 54 $\pm$ 8 degrees, confirming its slight elongation.

Table \ref{epoch} shows the results of fitting a two-dimensional Gaussian component to the emission. The consistent increase in the Gaussian flux peak with larger beam areas suggests that $\sim$15\% of the flux is on scales between 0”.1 and 1”. However, there are also indications of variability since the integrated flux density over the Gaussian at 150 mas resolution of the 2016 data is 17.89 mJy while in 2019, at 22 mas resolution, the integrated flux density is 21.35 mJy, amounting to a $\sim$20\% increase. In order to allow for irregularities we also measured the total flux within the 3 $\sigma_{\rm rms}$ contour for the combined emission observed in 2016 (\citep{Homan2018b}, ACA+TM2+TM1) and the new TM0 data. The 2016 (lower resolution) total flux was 19.3 mJy within 0.7'' from the peak, whilst the 2019 total flux is 24.6 mJy within 0.1'' from the peak. This is again a $\sim$20\% increase, greater than the expected combined flux scale and stochiastic uncertainties of $\sim$10\%.

Assuming that the star has a brightness of 4800 L$_{\odot}$ and a temperature of 3236 K \citep{Winters2007}, \citet{Homan2018b} estimated the continuum stellar flux contribution (by assuming black-body emission in a bandwidth 1.6 GHz around the rest frequency of 230.5 GHz) to be 22.7 mJy. It is known that the observable extent of an AGB star increases with wavelength \citep{Reid1997,Matthews2018}. So, for EP Aqr, we can assume photospheric radius $R_{\star}$ of 166 R$_{\odot}$, or about 6 mas at 135 pc, as estimated by \citet{Dumm1998}. The 230-GHz optically thick surface is unlikely to be more than $\sim$2$R_{\star}$ \citep[e.g.][]{Vlemmings2019} so even with uncertainties it has an upper limit of about 20 mas radius. Therefore, it seems likely that the radio atmosphere of the star in fact contributes only 15--18 mJy, hence additional, sporadic contributions to emission on larger scales are implied. These may be associated with companion interaction and/or a specific dust production events, and/or to variability of the radio atmosphere.



\begin{figure*}[]
        \centering
        \includegraphics[width=18cm]{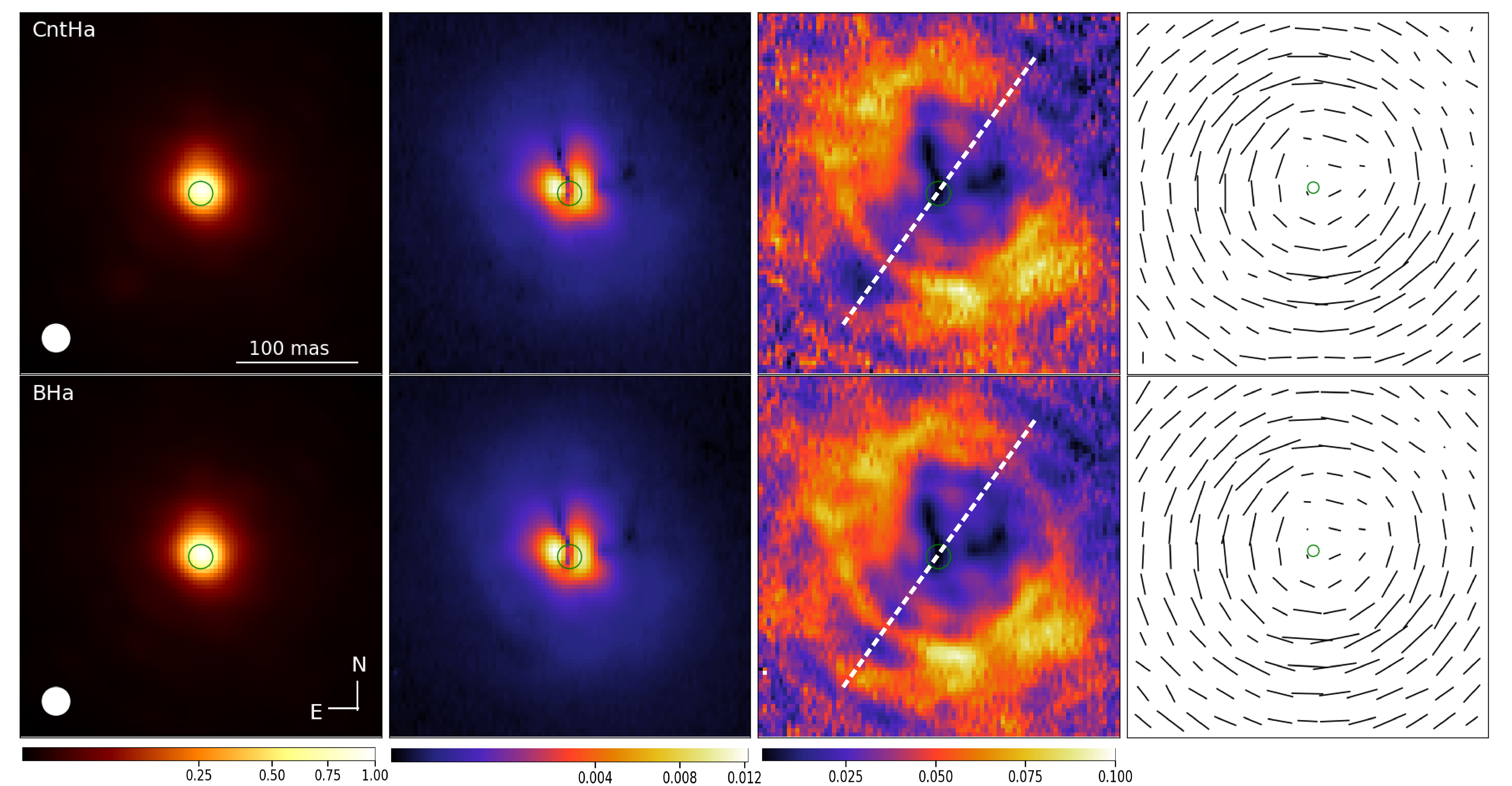}
        \caption{Overview of the SPHERE/ZIMPOL observations of the wind of EP Aqr. The top row represents the CntH$\alpha$ filter, the bottom row the BH$\alpha$ filter. The first column is the total observed intensity, the second column is the polarised flux, the third column is the degree of linear polarisation (DoLP), and the fourth column is the orientation of the polarisation vector. The beam size is shown in the bottom left of the leftmost panels. Both the intensity and polarised flux were normalized to the maximum intensity of each filter, thus preserving the degree of linear polarisation. The green circle in the centre represents the radius of the stellar photosphere as determined by the Stefan-Boltzmann relation. The white dashed line in the DoLP panels represents the orientation of the symmetry axis of the system as determined by \citet{Homan2018b}.
        \label{sphere}} 
\end{figure*}


\subsection{SPHERE observations} \label{spherecont}

Fig. \ref{sphere} shows the results of the unsaturated SPHERE/ZIMPOL observation of the inner wind of EP Aqr at length-scales in the order of 0.1''. The intensity images, which have a resolution of $\sim$27 milliarcseconds, exhibit a strong centrally condensed signal at the stellar position. The linearly polarised flux map appears similar, though it possesses some instrumental/reduction artefacts that split the otherwise predominantly round central feature in half vertically. This central artefact is seen on all ZIMPOL polarimetric images when the star is not resolved, including the PSF (See Fig .\ref{calib}). It does not question the real signal. The maps reveal two patches with an elevated degree of linearly polarisation localised to the north-west and the south-east of the stellar position. The presence of instrumental/reduction artefacts in the polarised flux maps and in the maps of the degree of linear polarisation (DoLP) suggests that the central portion of the data cannot be trusted. These bright DoLP regions, which are quite similar in shape and amplitude, exhibit maxima around $\sim$15\% of the total intensity, but are on average closer to $\sim$9\% of the total captured light. Furthermore, they appear to trace sections of a ring centred on the stellar position, but decrease in intensity and width as they approach the north-west to south-east axis. At their widest point, they measure about 0.09'', and both exhibit some substructure in the form of slight local DoLP enhancements of the order of a few percent, and with typical angular sizes of 0.05''. Despite these small variations, the overall distribution appears predominently smooth, and not clumpy. The locations where the DoLP drops down to the detection limits appear to be well aligned with the orientation of the symmetry axis of the wind of EP Aqr as determined by \citet{Homan2018b}, and which is overplotted on the DoLP plots in Fig. \ref{sphere} as a white dashed line. We elaborate on this coincidence in Sect. \ref{spherediscuss} in the discussion. 


\begin{center}
\begin{figure*}[]
        \includegraphics[width=17cm]{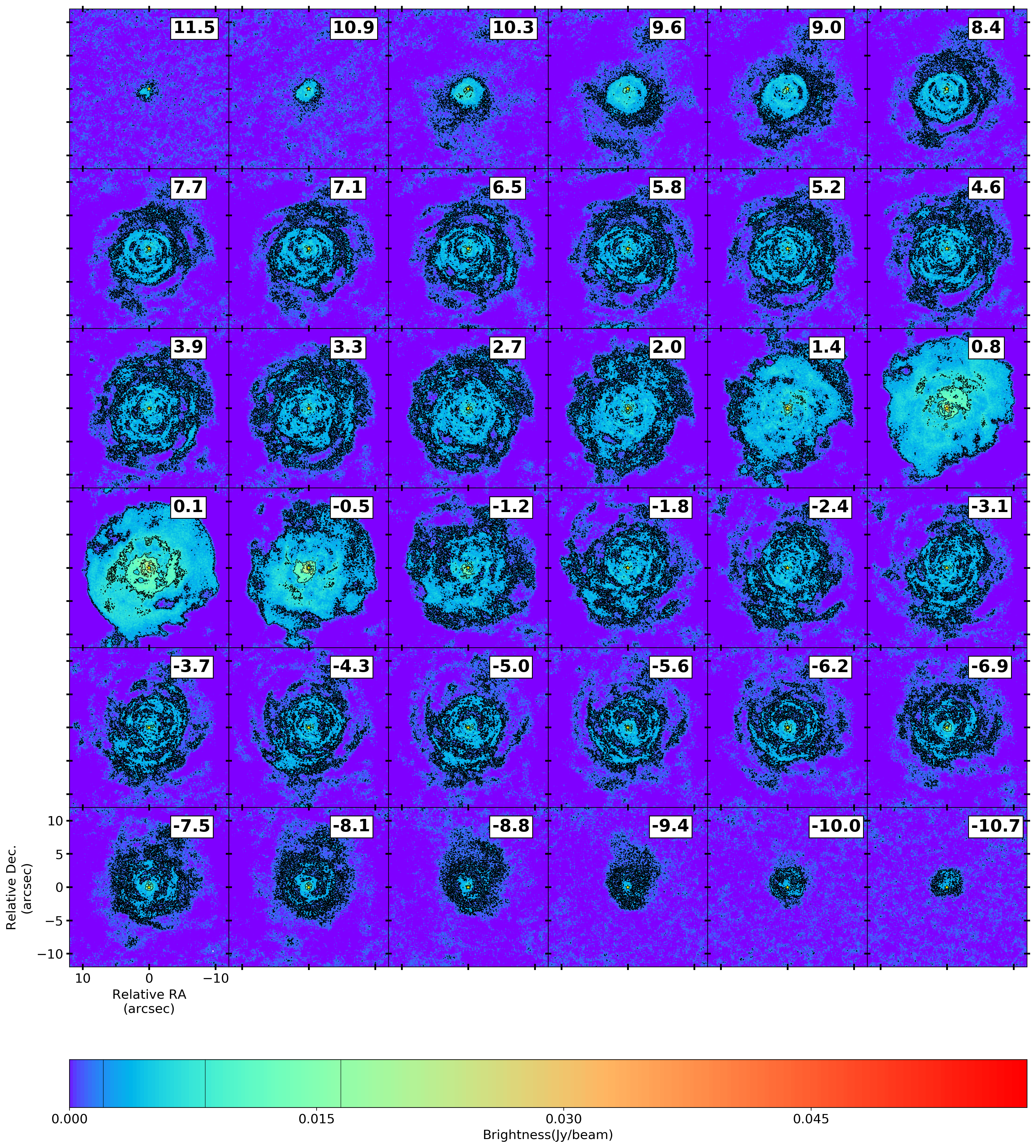}
        \caption{Continuum-subtracted channel maps of the $^{\rm 12}$CO emission. The labelled velocities have been corrected for $v_{*}$. The central observed frequency is the average over the observation epochs. The contours are drawn at 3, 12, and 24 times the rms noise value outside the line ($\sigma_{rms}$ = 6.86$\times {\rm 10}^{\rm -4}$ Jy/beam). Length scales are indicated in the bottom left panel. The maps are centred on the continuum peak position, which is indicated by the yellow star symbol. The beam size is illustrated in the bottom-right corner of the bottom-left panel (0.087''$\times$0.079'').
        \label{COchan}} 
\end{figure*}
\end{center}

\begin{figure*}[]
        \centering
        \includegraphics[width=16cm]{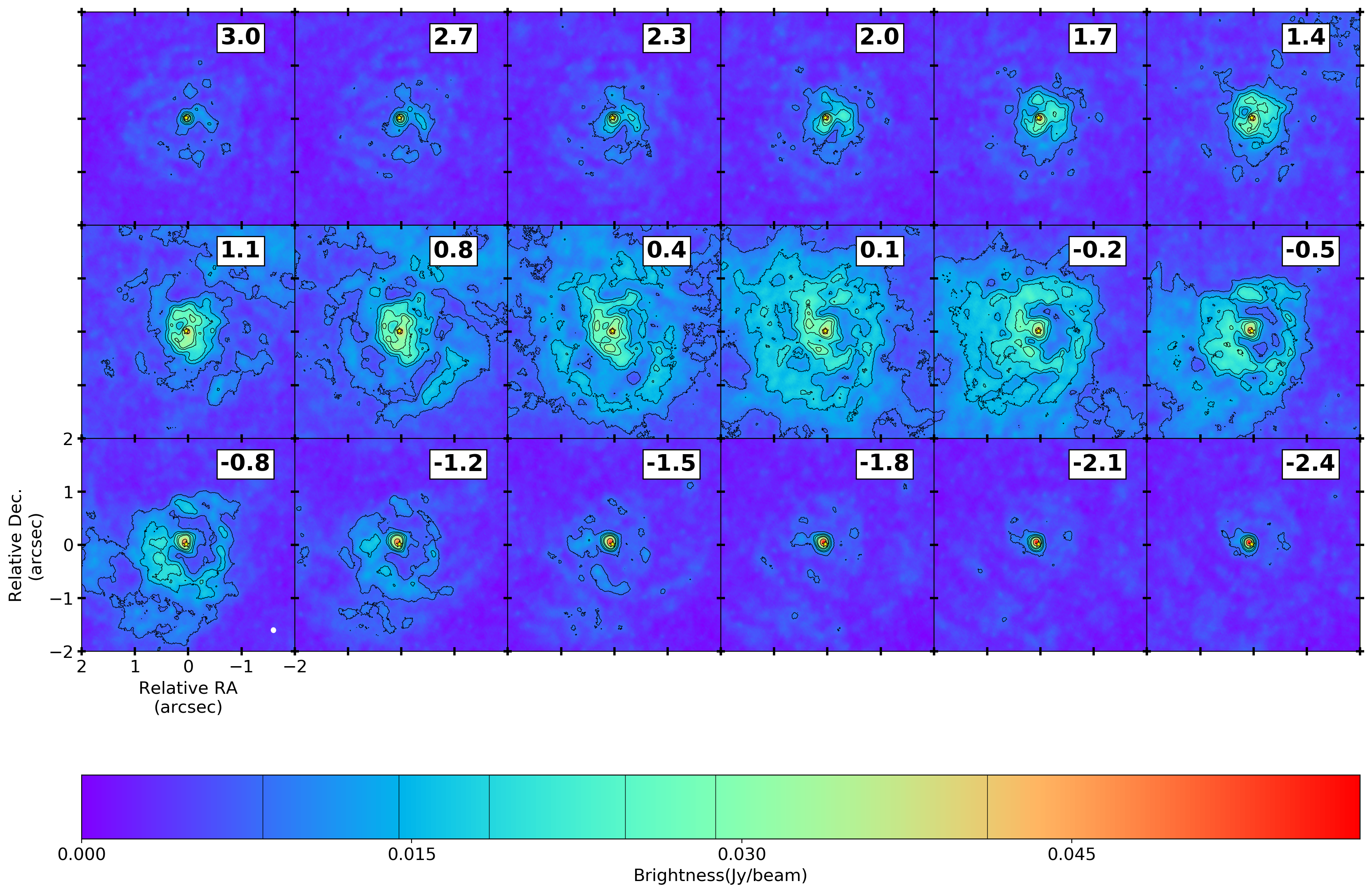}
        \caption{Identical channel maps to Fig. \ref{COchan}, but with contours drawn at 12, 18, 24, 30, 36, 42, and 60 times the rms noise value outside the line ($\sigma_{rms}$ = 6.86$\times {\rm 10}^{\rm -4}$ Jy/beam, and zoomed in on the central 4''$\times$4'' to highlight the complexity of the inner wind region.
        \label{COcz}} 
\end{figure*}

\begin{figure*}[]
        \centering
        \includegraphics[width=16cm]{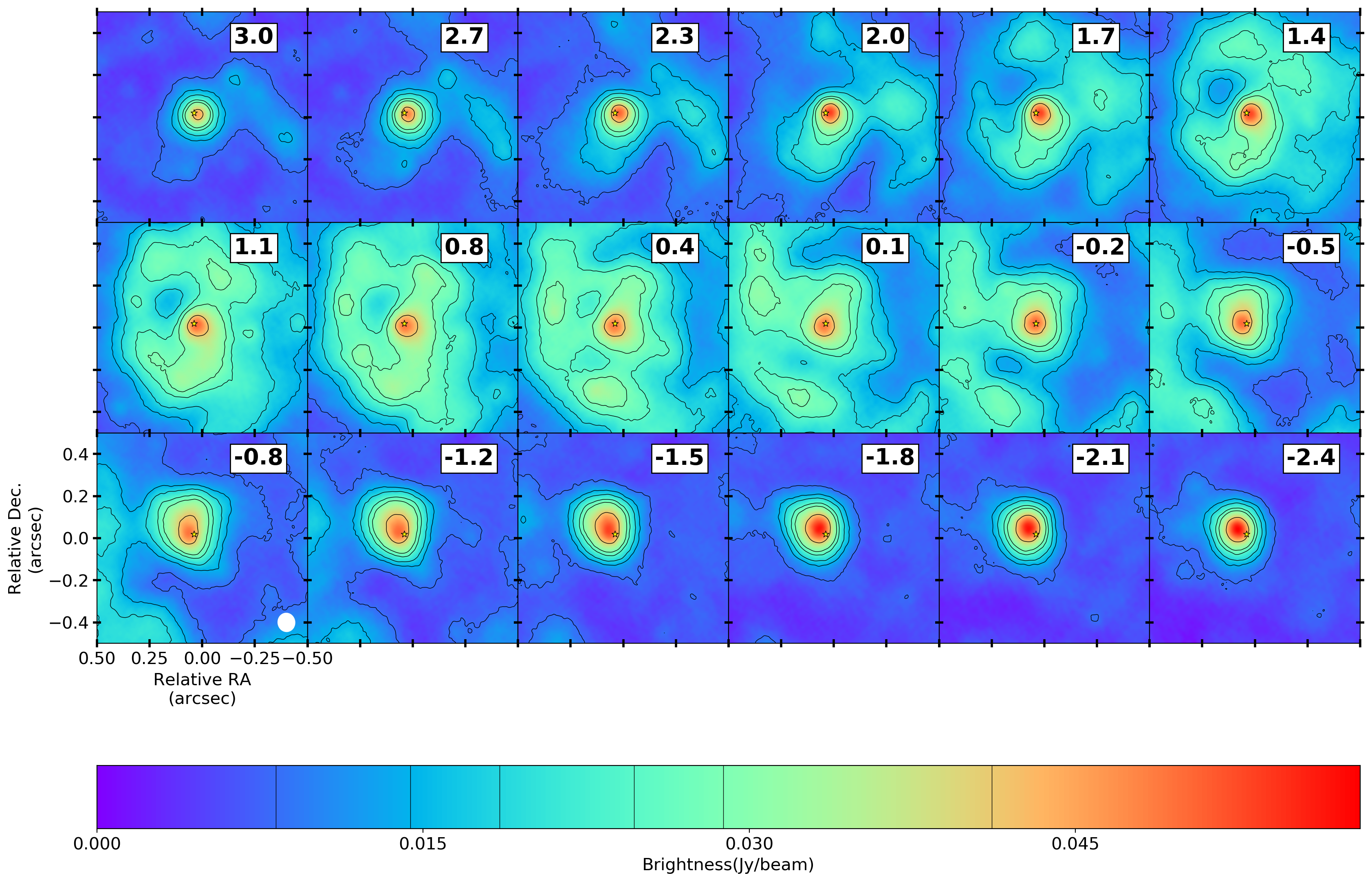}
        \caption{Identical to Fig \ref{COcz}, but zoomed in on the central 1''$\times$1'' to highlight the position change of the CO emission brightness peak with respect to the continuum brightness peak position (star symbol).
        \label{COcsz}} 
\end{figure*}

\begin{figure}[]
        \centering
        \includegraphics[width=8.5cm]{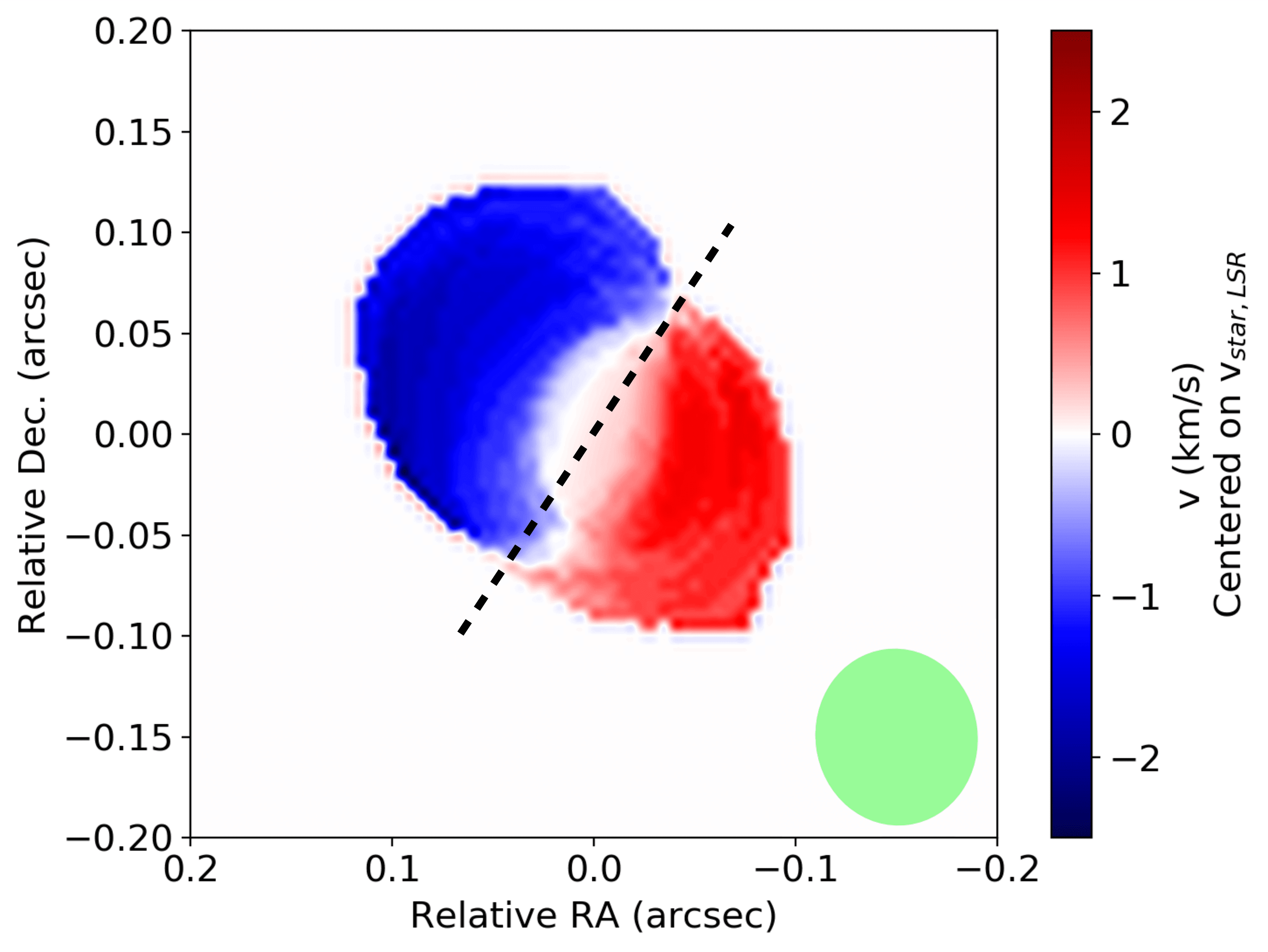}
        \caption{Moment 1 map of the inner CO emission for the -3$\kms$ to 3$\kms$ velocity range, and centred on the continuum brightness peak position. Only the signal above 24 times the noise rms is shown. The beam is shown in green in the bottom right corner. The black dashed line shows the approximate orientation of the rotation axis, with a PA of 145$^\circ$.
        \label{COmom1}} 
\end{figure}

\section{ALMA molecular emission} \label{molec}

In this section we describe the observational results for the molecular lines for which we successfully combined the 2016 data with the latest observations. The 2016 data did not contain the $^{\rm 13}$CO and SO$_{\rm 2}$ spectral lines observed in the 2019 campaign, so we do not include these in this article for the sake of consistency. All mentions of blue-shift and red-shift are with respect to the stellar velocity, which is assumed to be $v_{\rm lsr} = -34$\,km\,s$^{-1}$. For optimal appraisal of the fine detail in the data figures we recommend that all images presented in this paper are displayed on screen.

\subsection{$^{\rm 12}$CO $J=\ $3$-$2 emission} \label{COobs}

In this section we describe the nature of the $^{\rm 12}$CO $J=\ $3$-$2 emission. In addition to the initial analyses performed by \citet{Homan2018b}, the increased angular resolution allows for a better visualisation of the finer details in the observed emission patterns. Fig. \ref{COchan} shows the line channel maps of the combined data, and Figs. \ref{COline} and \ref{COmom0} in the appendix show the spectral line and the moment 0 map of the combined data, respectively.

Previous work by \citet{Homan2018b,Hoai2019} and \citet{TuanAnh2019} has shown that the emission features in the high-resolution interferometric data on EP Aqr are caused by a face-on, dense equatorial density enhancement (EDE), symmetrically encased by two separate wind hemispheres, which are the sources of the central peak and the broad plateau in the spectral line, respectively. The dense EDE was shown to harbour a counter-clockwise oriented spiral shape, and its narrow spectral extent of $\sim$3$\kms$ was interpreted as indication that its thickness was quite limited. In the centre of the EDE, they detected highly complex emission pattern at the base of the spiral. The channel maps, zoomed-in on the inner 4''$\times$4'' of the $^{\rm 12}$CO emission, and spectrally confined to the central peak in the emission line (Fig. \ref{COline}), are shown in Fig. \ref{COcz}. At the stellar velocity a large, nearly complete ring of emission can be seen surrounding the continuum brightness peak (CPB) position. The CBP position is not in the centre of the circle, but rather to its west. To the east of the CBP position, a central arc can be seen extending from north to south, which grazes the CBP position, and appears to divide the circular feature into a western and an eastern half. Due to the curved nature of this central arc and the presence of the bright central emission centred on the CBP position, the low-emission parts appear as crescent-shaped, one larger one to the east of the CBP position, and a smaller one to the west. The correlation between the central, northwards curving arc, and the larger circular feature could be interpreted as the first winding of a clockwise spiral. However, this is opposite to the orientation of the larger spiral that \citet{Homan2018b} discovered in the EDE, and can therefore not straightforwardly be recognised as the base of the spiral in the EDE. The overall shape of the emission in EP Aqr's inner wind was tentatively interpreted by \citet{Homan2018b} as the spectral manifestation of morphology associated with the wind-Roche-lobe-overflow scenario proposed by \citet{Mohamed2012}, and is visualised in their manuscript's Fig. 3, though their image must be flipped vertically to be comparable with the situation in EP Aqr's wind.

With the combined datacubes, the emission distribution of the inner part of the EDE of EP Aqr is much better resolved, and can thus be investigated in much finer detail. The large, nearly circular arc covering the south, east, and part of the north of the central channel maps can now be seen to be composed of a smooth background covered in higher-emission clumps with sizes up to a few beams at most. It has a radius of approximately 1.25'', and its width, measured from the 18$\times \sigma_{\rm rms}$ contour, varies between 0.3'' to 0.45''. Furthermore, the eastern crescent appears to be quite perturbed, and actually not very crescent-shaped at all. Only the western crescent-shaped void, which is only visible in the blue-shifted channels, remains distinctly recognisable. Its axis of symmetry lies east-to-west, and on this axis the void is centred on a position located approximately 0.4'' to the west of the CBP. The void is present in all blue-shifted channels between 0 and -1.5 $\kms$, beyond which it blends with it surroundings. This velocity width corresponds exactly with the spectral half-width of the EDE.

The red-shifted channels (1 to 3 $\kms$ range) reveal an interesting small emission feature to the north and west of the CBP. At 3 -- 2$\kms$, this feature is clustered predominantly around 0.4'' to the west of the CBP position. As the channel velocity gradually progresses towards the stellar velocity, the emission cluster becomes more elongated, and traces an arc-like course over the north of the CBP position, and to the east. And though barely resolved, this arc remains clearly distinct from the brightness peak of the CO maps. Then, around 1 $\kms$, the arc blends with the additional emission features that start to dominate the inner wind of EP Aqr. Both the blue-shifted and red-shifted emission show some very interesting behaviour around 0.4'' to the west of the CBP position, which coincides with the location of the significant emission void in the SiO emission discovered by \citet{Homan2018b}. We elaborate on the peculiarity of this position in Sect. \ref{companion} in the discussion.

The brightness peak of the CO emission can be seen to shift position with respect to the CBP position as a function of velocity, as is illustrated in Fig. \ref{COcsz}. For the most red-shifted velocities ($\sim$3$\kms$), the CO brightness peak is located approximately 2 mas to the west of the CBP position. and it gradually shifts eastwards as velocities become more blue-shifted. Around stellar velocity it centres on the CBP position, and at higher blue-shifted speeds ($\sim$-2$\kms$) the CO brightness peak is distinctly positioned 5 mas to the east of the CBP position. This spectral progression is precisely in agreement with the findings of \citet{Homan2018b} who discovered what appeared to be gas in (rigid) rotation from their analysis of the SO$_{\rm 2}$ emission within a radius of 2''. In Fig. \ref{COmom1} the position shift of the CO brightness peak as a function of projected speed is captured in a moment 1 map. The map is indeed comparable to the moment 1 map of SO$_{\rm 2}$ revealed by \citet{Homan2018b}, but due to the increased resolution it can be seen to be a factor 2 more compact. The white zero-velocity transition zone between the blue- and red-shifted emission, which has been interpreted as the orientation of the rotation axis of the gas in the plane of the sky, has a PA of $\sim$145$^\circ$. 

These data also highlight the large amount of sub-structure that is present in the blue-shifted and red-shifted outflow hemispheres of this AGB stellar wind, which is manifested by the wings of the spectral line. The structure observed here consist of distinct filaments, which appear to trace segments of nearly complete circular shells in the maps. An example of such a nearly complete circle can be seen around velocities of 5.2--5.8$\kms$ (see Fig. \ref{COchan}). And while not perfectly identical, such structures can clearly be recognised both in the red-shifted and the blue-shifted line wing, and have smaller radii at larger projected velocities.

As was first noted by \citet{Homan2018b}, these arcs exhibit a slight deviation from concentricity, with the red-shifted emission pulling slightly to the south-east, and the blue-shifted emission to the north-west. The arcs in the blue- and red-shifted line wings also differ quite substantially in number and size. Though generally difficult to differentiate from each other due to the weak, clumpy emission, we have counted them by eye by summing up the generally circular features that appear in the emission. At the 3 times $\sigma_{\rm rms}$ level only one shell-like feature can be discerned. The other shells appear in higher signal-to-noise regions, and must therefore be appraised at the 6 times the $\sigma_{\rm rms}$ level. These contours reveal 6 distinct shells for the red-shifted line wing (See Fig. \ref{COchan84} in the Appendix). In totality, the outflow hemispheres thus exhibit 7 separate shells in the part of the outflow that is moving away. These arcs have typical widths that increase with radius. Assuming a constant velocity of 11 $\kms$ in this part of the wind, we can calculate the deprojected radii of the most obvious shells in the data. We found arc widths of approximately 1.5'' for the largest ring (at 3$\sigma_{\rm rms}$ level) located at a distance of 7'', of 0.4'' for rings with a radius of 3'', down to 0.3'' for rings with a radius of 1.2''. Though the arcs have an irregular outline at the enclosing contour level, the distance between the arcs appears to vary between 0.5'' and 0.8'', except for the two outermost ones, which appear somewhat more detached ($>$ 2'') form the bulk of the emission. In the blue-shifted wings of the outflow the arcs are much more difficult to differentiate, with only 2 recognisable arc-like features at the 3$\sigma_{\rm rms}$ level, and no other clearly distinguishable arc-like features at higher rms noise levels. We comment on the peculiarity and potential origin of these arc-like features in Sect. \ref{comp2}.

\subsection{$^{\rm 28}$SiO $J=\ $5$-$4 emission} \label{SiOobs}

\begin{center}
\begin{figure*}[htp]
        \includegraphics[width=17cm]{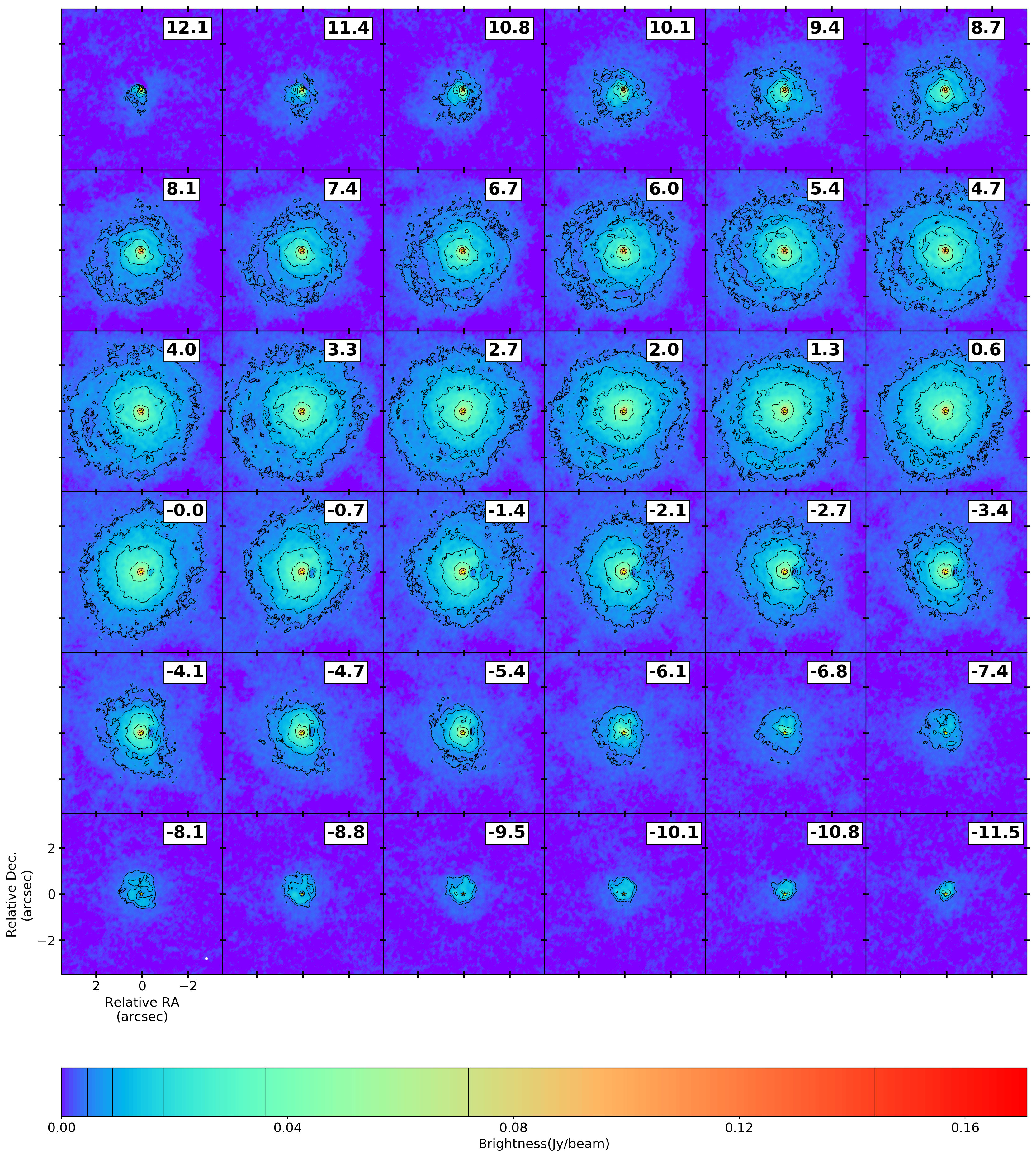}
        \caption{Continuum-subtracted channel maps of the SiO emission. . The labelled velocities have been corrected for $v_{*}$. The central observed frequency is the average over the observation epochs. The contours are drawn at 3, 6, 12, 24, 48, and 96 times the rms noise value outside the line ($\sigma_{rms}$ = 1.5$\times {\rm 10}^{\rm -3}$ Jy/beam). Length scales are indicated in the bottom left panel.  The maps are centred on the continuum peak position, which is indicated by the yellow star symbol. The beam size is illustrated in the bottom-right corner of the bottom-left panel (0.085''$\times$0.079'').
        \label{SiOchan}} 
\end{figure*}
\end{center} 

\begin{figure}[]
        \centering
        \includegraphics[width=7cm]{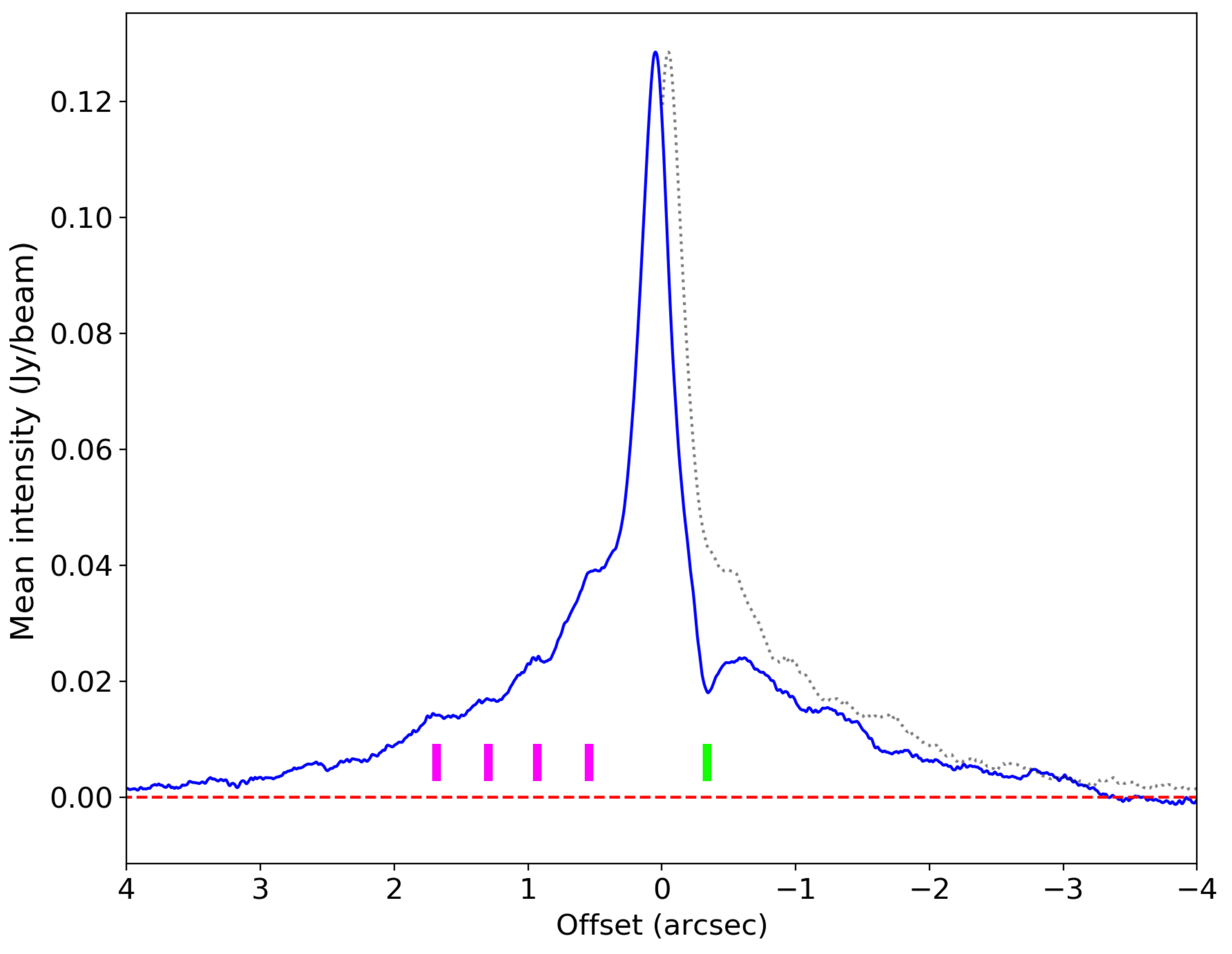}
        \includegraphics[width=7cm]{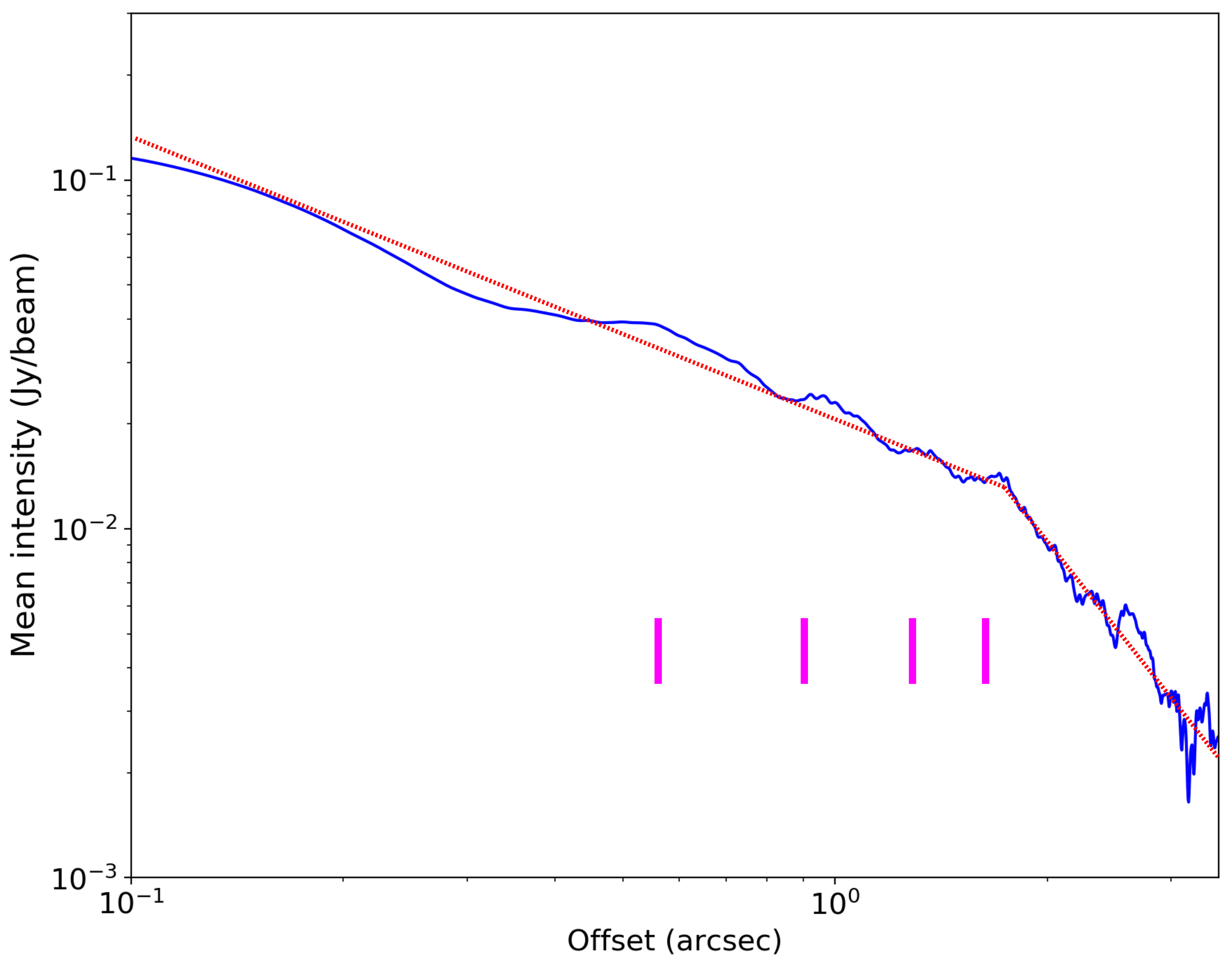}
        \caption{\emph{Top:} SiO emission profile of the central channel, extracted from east to west through the center. The intensity axis has units of Jy/beam. The eastern (left) profile is overplotted as a grey dashed line over the western (right) profiles to highlight the asymmetry. The SiO void is highlighted by the green tag. The purple tags indicate the locations of the equidistant bumps in the inner portion of the SiO profile. \emph{Bottom:} Eastern profile in log-scale. The overplotted red dotted line exhibits a shift in regime between a slope of -0.77 for the inner 1.8'', and -2.6 for the outer part. }
        \label{sioprofile} 
\end{figure}

The channel maps of the combined array data of the SiO emission can be found in Fig. \ref{SiOchan}. As discussed by \citet{Homan2018b}, the emission is comparatively smooth and spherical compared to the CO emission. This is supported by the shapes of the emission line and moment 0 map, shown in Figs. \ref{SiOline} and \ref{SiOmom0} in the appendix, respectively. This is to be expected for the SiO molecule, since it is highly abundant in the inner winds of AGB stars and has a strong dipole moment. This implies that the stellar infrared radiation dominates the molecular energy level excitation close to the star. Hence, contrary to the CO molecule that has a very low electric dipole moment, any collisional effects, which act as a proxy for local density, only have a limited influence on the population of the SiO energy levels with the result that density variations are not expected to be prevalent in SiO emission. The SiO spectral line does not exhibit any unusual features besides a slight but important asymmetry, which is caused by decreased emission in the negative portion of velocity-space. This can be attributed to the emission void first detected in \citet{Homan2018b}. The higher-resolution data now permit a better analysis of the properties of this void, for which we show zoomed-in channel maps in Fig. \ref{SiOcz} in the appendix.

The void, which is clearly resolved in the combined data, has an elongated shape, with a vertical (north-to-south) size of approximately 0.4'', and a horizontal size of approximately 0.2''. Its lowest emission point drops down well below the 3$\sigma_{rms}$ detection limit, and is located approximately 0.4'' directly west of the continuum peak position, which at a distance of 135 parsec corresponds to a physical distance of $\sim$55 AU. The void appears precisely at the stellar velocity in the EDE, and remains a dominant spectral feature in the data down to velocities of -6 $\kms$, which are well inside the wind hemispheres. Beyond this velocity it blends with the surrounding low emission. The deep penetration of this void into the blue-shifted wind hemisphere implies that the void is in fact more like a hollow tube, which originates in the EDE (see Sect. \ref{COobs}, and \citet{Homan2018b,Hoai2019} and \citet{TuanAnh2019}). This tube extends nearly perpendicular to the plane of the EDE, penetrating the blue-shifted portion of the outflow while not penetrating the red-shifted part at all. In combination with the suggestion by \citet{Homan2018b} that this void may be the manifestation of a local environment created by a photodissociating radiation field caused by the presence of a companion, this asymmetric tube-like behaviour yet again provides evidence that the outflow of EP Aqr consists of two wind hemispheres, separated by a dense and flat EDE. With these improved data we can use the size of the void to estimate the local conditions needed to explain the observations. We defer further discussion to Sect. \ref{companion}.

These new data also enable the identification of some structures that were not detected in the 2016 data. In particular, the red-shifted channels around 6$\kms$ reveal a distinct arc or outline of a shell-segment to the east of the emission zone, at the level of the 3$\sigma_{rms}$ contour. The shell-segment has a radius of approximately 2.6'', a thickness of approximately 0.6'', and extends most notably to the east and south-east. As mentioned above, detecting density variations so clearly is quite peculiar considering the dominant excitation mechanism for SiO is IR pumping. It is unclear why this particular arc emerges from the SiO emission.

For the sake of completion we show the position-velocity diagram of the SiO emission along the PA = 150$^\circ$ axis in Fig. \ref{SiOpvd}, constructed using slit width equal to the beam size. It possesses the typical shape associated with spherical emission in a predominantly radial outflow. Thanks to the high-resolution of the data, even the blue wing absorption over the stellar surface van be discerned. However, of main interest is that the data clearly shows an extension beyond what is typically associated with the wind speed of EP Aqr, as first discovered by \citet{Homan2018b}. These features are asymmetric in velocity-space, we measure their maxima to extend to approximately -18 and +22 $\kms$. In later work, \citet{TuanAnh2019} interpreted these feautres as narrow high-velocity polar streams of gas that originate from within a radius of 25 AU from the stellar source. If so, then these polar streams could be generated by a tertiary companion that resides within 10 AU from the AGB star. We elaborate on this thought in Sect. \ref{comp2}.

In addition, even though the SiO emission distribution appears relatively smooth in the channel maps, the radial emission profile shown in Fig. \ref{sioprofile} reveals that in fact the SiO emission does possess some low-amplitude substructure. In this figure, we show how the emission varies on an east-to-west oriented axis with a width of one beam, and the emission is averaged over the beam width. The axis runs through the CBP position, for the channel at stellar velocity. This is the channel through the bulk of the EDE surrounding EP Aqr, and is also the first channel to exhibit the void that appears in the right (western) side of the profile, highlighted by a green tag. The emission on this side is clearly weaker than its eastern counterpart, as can be seen in the difference between the blue and grey dashed profiles in Fig. \ref{sioprofile}. The eastern profile, which is presumably the least affected by the presence of the void, shows monotonically decreasing emission following what appears to be two distinct regimes. The emission in the inner portion of the wind can be approximated by a power-law with a slope of -0.77.  Around 1.8'', this regime abruptly shifts to a much steeper radial slope, which can be approximated by a power-law with a slope of -2.6. This behaviour can be reasonably attributed to a radiative transfer effect, with the outer regions probing optically thin SiO emission, and the slope shift representing the transition towards optically thick SiO emission at smaller radii. Under this assumption, we can to first order state that if the SiO abundance and level excitation is approximately constant within this region, the density drops on average as $r^{\rm -2.6}$ in the EDE of EP Aqr. Detailed three-dimensional radiative transfer modelling of discs around AGB stars has shown that such steep density profiles could be typical in the context of AGB EDE's \citep{Homan2017, Homan2018}. However, these objects were found to be very compact and to show signs of differential rotation. They are therefore quite different from the EDE in the wind of EP Aqr. Thus, we are not certain whether the same conclusions apply in this context. At this point, however, we cannot conceive any other mechanisms that would explain such a steep slope.

Finally, we note the presence of 4 small consecutive bumps that occur roughly every 0.45'' present a first order deviation from a smoothly decaying profile. These bumps do not have an equivalent counterpart in the CO emission (see Fig. \ref{COprofile} in the appendix). We elaborate on what may be the origin these bumps in Sect. \ref{comp2}


\section{Discussion} \label{discus}

\subsection{Evidence for companion in the wind} \label{companion}

The strongest evidence for the presence of a companion is the overall morphology of the wind. As has been extensively analysed by \citet{Homan2018b} and \citet{Hoai2019}, the resolved emission patterns can be best explained by a nearly face-on, spiral harbouring equatorial density enhancement (EDE), encased on both sides by less dense wind hemispheres that could be considered a bi-conical outflow. For slowly rotating giants such as AGB stars, such global features are impossible to explain without invoking the presence of a companion. This companion will bend the otherwise radial wind streaming lines towards the orbital plane of the system, generating an equatorial overdense region with an additional narrow spiral generated from its gravity wake \citep{Kim2011,Liu2017}. As such, the global morphology of the wind is shaped. There is, however, additional evidence in the observed emission patterns that a companion is present in the wind.

Further evidence can be found at the base of the spiral, which shows a distinct hole in the emission that is especially pronounced in SiO. This so-called emission void, which is located about 0.4'' to the west of the continuum brightness peak position, was first discovered by \citet{Homan2018b}, and we describe its properties in Sect. \ref{SiOobs}. For a CO inter-spiral-arm distance of about 3'' - 4'', the dynamical crossing time is $\sim$200-250 years for an assumed wind speed of $\sim$10$\kms$. This crossing time corresponds with the orbital period of an object located at approximately 50 AU, or 0.37'' from the central star. Considering the uncertainty on the wind speed in EP Aqr's EDE, which so far has not been properly measured due to its nearly face-on orientation, this back-of-the-envelope calculation shows the striking similarity between the derived location of the tentative companion and the position of the SiO emission void.

\begin{figure}[]
        \centering
        \includegraphics[width=8.5cm]{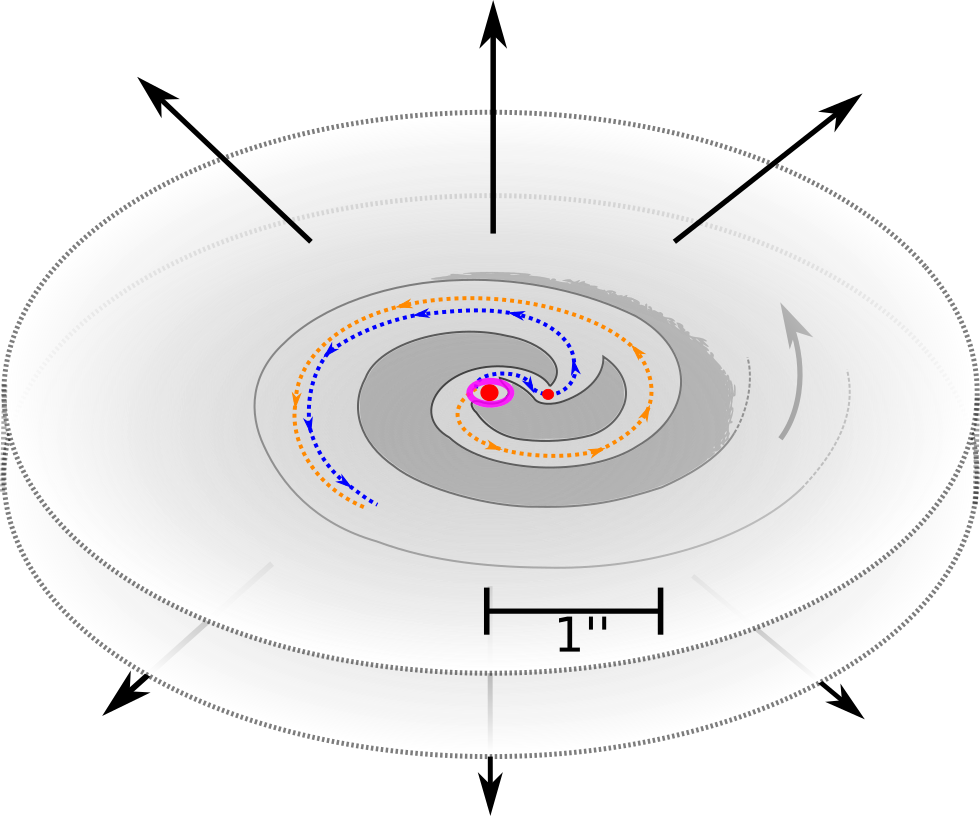}
        \caption{Schematic diagram of the wind-companion interaction zone. The two red dots represent the AGB star and the companion, separated by 0.4''. The pale outline in the central grey (less dense) zone represents the local density enhancement as a consequence of the interaction hydrodynamics. The blue (orange) dotted arrow line represents the flow along L1 (L3), which combine into the larger scale spiral observed in the EDE. The three arrows on top and bottom represent the (bi-conical) AGB outflow. We have indicated the dust ring observed with SPHERE/ZIMPOL in fuchsia.
        \label{diagram}} 
\end{figure}

Additionally, close inspection of the distribution of CO emission (whose distribution predominantly follows the wind density) in the inner wind of EP Aqr reveals a clear filament of gas originating a the location of the SiO void at $\sim$3 $\kms$, and curving northwards over the CBP position at lower speeds (see yellow arc in Fig. \ref{COhole} in the appendix). This feature was first detected by \citet{Homan2018b}, but as rightly pointed out by \citet{TuanAnh2019}, an error was made in the sign of the considered velocity-channel. This CO emission filament appears only in the red-shifted channels, as described in sect. \ref{COobs}. Comparison with hydrodynamical models such as the ones by \citet{Mohamed2012} and \citet{Liu2017} reveals that, for companion masses comparable to the mass-losing star's, this could be the formation of the spiral shock as a result of interaction with the companion. These hydrodynamical models predict that a large portion of the matter lost by the AGB star leaves the system in two distinct streams that flow along the Lagrange points L1 and L3. The stream along L3 is typically associated with reflex motion of the mass-losing AGB star around the centre of mass of the binary system (henceforth \emph{reflex-spiral}), while the flow along L1 is caused by the direct attraction of the outflow material by the companion\citep{Liu2017}. The latter is launched into the CSE at augmented speeds after its closest interaction with the companion, and creates what is typically referred to as the \emph{wake-spiral}. Considering the orientation of the larger-scale CO spiral discussed by \citet{Homan2018b}, the companion must be rotating around the AGB star in a clockwise fashion. This implies that the position of the filament and its arc-like spectral evolution around the CBP position, while remaining in the red-shifted portion of the cube, indeed fit the dynamics of the gas that probes the formation of the wake-spiral. By extension of the argument, the filament that defines the southern edge of the western crescent could then be the reflex-spiral (see red arc in Fig. \ref{COhole} in the appendix). We present a schematic diagram of the proposed gas flow pattern in the inner wind of EP Aqr in Fig. \ref{diagram}. These features, which emerge from the higher-resolution data, strengthen \citet{Homan2018b}'s initial hypothesis that the detected shapes are associated with the hydrodynamical wind shaping caused by the presence of a companion at 0.4'' to the west of the CBP position.

There exists a spatial and spectral coincidence between the SiO emission void (see Fig. \ref{SiOchan}) and a similar, but spatially somewhat larger crescent-shaped void in CO (see Fig. \ref{COhole} in the appendix). As mentioned in Sect. \ref{COobs}, \citet{Homan2018b} suggested that the crescent shaped decreased emission zones could be the morphological manifestation of wind-Roche-lobe-overflow (WRLOF) \citep{Mohamed2012}, which pointed them towards invoking the presence of a companion. The current data reveals that the location of the centre of the crescent shape in CO and the centre of the void in SiO are approximately equal (see Fig. \ref{COhole}). This could suggest that (1) both the void in SiO and the crescent in CO may have the same origin, and/or (2) that the WRLOF scenario alone may not the best model to explain EP Aqr's inner wind morphology. This could imply that the observed shapes may be the result of a different type of wind-companion hydrodynamical interaction, combined with emission-dampening physics such as cooling or photodissociation. It must be noted, however, that there are similar, albeit smaller, emission gaps between arcs at many PAs in the inner wind region in the -2 to 2 $\kms$ velocity range. Hence, the discussed crescent-shaped void could also simply be yet another such gap between the complex sequence of arcs and clumps, and its coincidence with the SiO gap merely coincidental.

\subsection{The nature of the companion} \label{wd}

In this section we will estimate the properties of the companion under the assumption that the observed emission hole in SiO is a consequence of photodissociation. \citet{Homan2018b} estimated that if the companion was a main-sequence star, then its non-detection in the continuum yielded an upper limit on its mass of 0.1$\mso$. However, it is questionable whether such an M6 dwarf can generate the required pull on the gas to shape the wind to the extent that is observed \citep{ElMellah2020}. Furthermore, with a typical temperature of 2800K and luminosity of $\sim$10$^{\rm -3}\lso$ \citep{Siess2000}, its capability to sustain a flux of sufficient UV photons to explain the emission void observed in SiO and CO, either photospherically or via the creation of a hot accretion disc, is unlikely. 

With the current high-resolution data the SiO hole is resolved, and hence we can estimate the size and temperature of the photodissociating blackbody by relating the predicted UV photodissociation time scale with the dynamical time scale associated with the wind of EP Aqr crossing the SiO emission void. Assuming the companion is in the centre of the void, half of the void's width ($\sim$27 AU/2) divided by the wind speed ($\sim$11$\kms$) corresponds to approximately 5.8 years. 

\begin{figure}[]
        \centering
        \includegraphics[width=8.5cm]{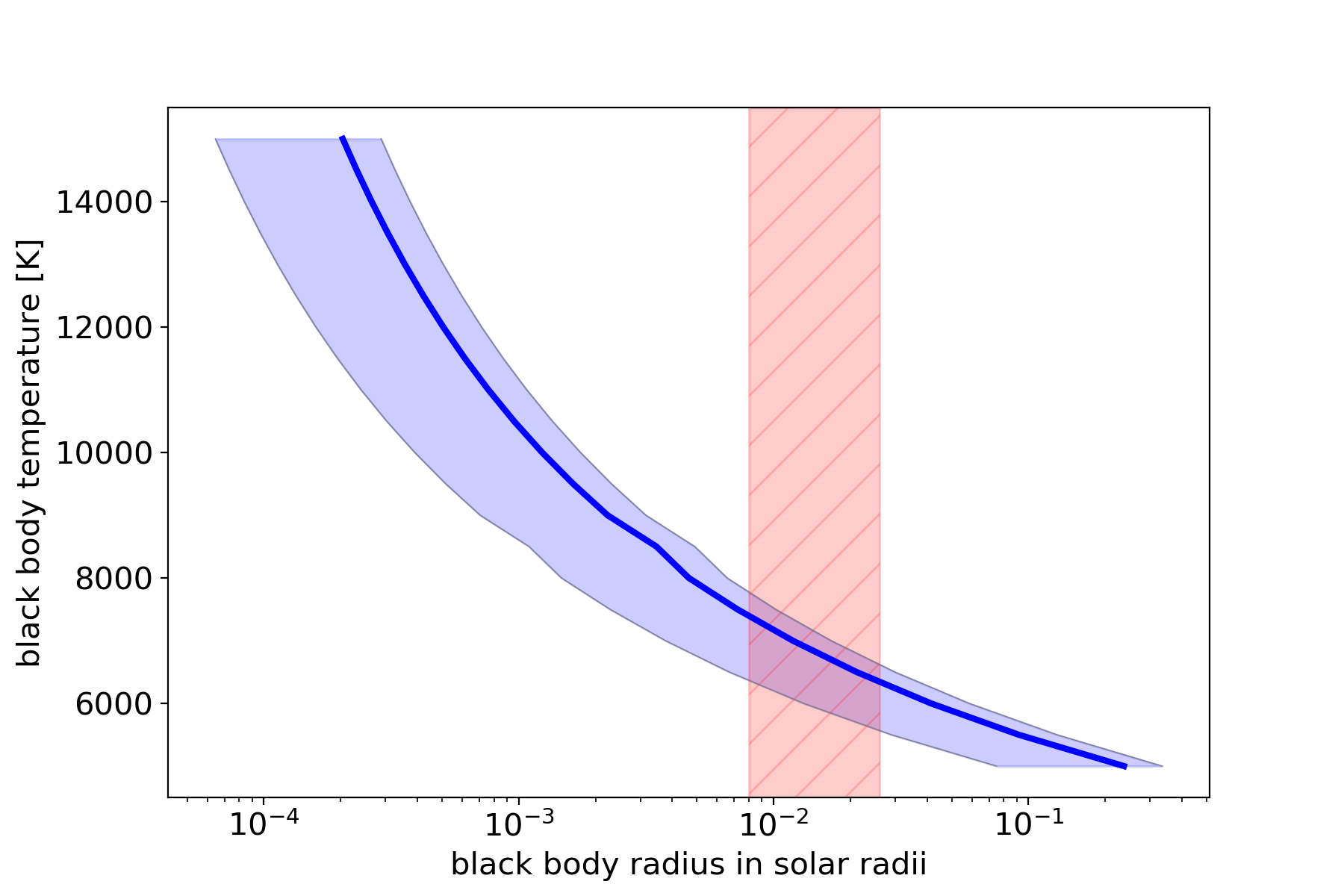}
        \caption{Blackbody temperature and radius combinations that would open-up a hole of the observed size by photodissociation of SiO. The blue curve is calculated for a wind speed of 11$\kms$. The translucent blue zone represents the space of solutions for a maximum deviation of 90\% on the assumed wind speed. The red hatched zone represents the white dwarf radii found in the literature.}
        \label{pd} 
\end{figure}

In order to calculate the photodissociation rate of SiO, we have made use of the cross-sections made available via the Leiden database for photodissociation \footnote{https://home.strw.leidenuniv.nl/~ewine/photo/index.html} which shows that it is photodissociated through five absorption lines at wavelengths between 100 and 140 nm \citep{Heays2017}. Using this data, we have calculated the UV photodissociation rate $\beta(r_{sc})$ of SiO for 21 evenly spaced black body temperatures between 5000K and 15000K, at an arbitrary scaling radius of $r_{sc}$ which is assumed to be 50 times the radius of the black body. The results of this calculation can be found in Table \ref{pdtable} in the appendix. The actual photodissociation rate at a distance $r$ from the black body can then be calculated by accounting for geometric dilution and extinction through dust, captured in the following expression

\begin{equation}
 \beta(r) = \beta(r_{sc})\left(\frac{r_{sc}}{r}\right)^2{\rm exp}(-\kappa_d \rho_d r),
\end{equation}

where $\kappa_d$ is the dust opacity and $\rho_d$ the dust density at $r$. These will be assumed to both be constant, and have values of $\kappa_d$=7cm$^2$/g \citep{Demyk2017} and $\rho_d$ equal to the gas density $\rho_g$ times an assumed dust-to-gas mass ratio of 200. For this exercise, $\rho_g$ is calculated using mass conservation. However, because we currently have no proper constraints on the density contrast between the EDE and the outflow, we use the ratio of the integrated fluxes between the CO line peak and the plateau as proxy for the density contrast between the EDE and the outflow hemispheres. This is an absolute lower limit on the actual value, but considering the unknowns this is currently the best that can be done. The photodissociation time scale is then given by $\beta(r)^{-1}$. Setting this photodissociation time scale equal to the dynamical void crossing time scale produces a space of solutions for black body temperature and radius combinations. The solution curve is visualised in Fig. \ref{pd}.

\begin{figure}[]
        \centering
        \includegraphics[width=8.5cm]{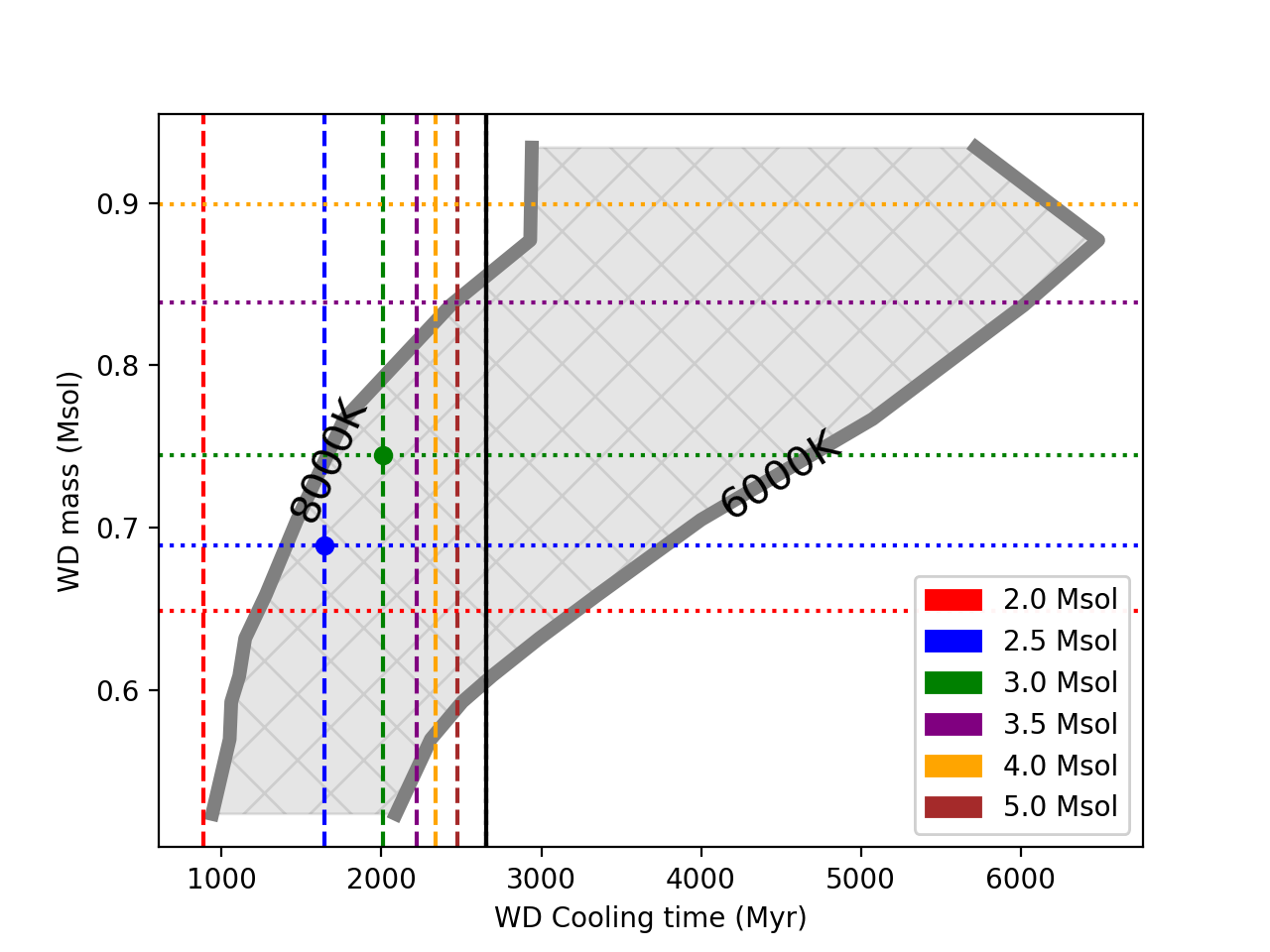}
        \caption{WD mass as a function of WD cooling time. The hatched region between the two broad grey curves represents the range of WD temperatures as derived from Fig. \ref{pd}. The black vertical line is the main sequence lifetime of a 1.7$\mso$ star, and therefore the WD cooling time upper limit. The zone to the left of the vertical dashed lines are the maximal cooling times for WD's with respective colour-coded progenitor masses. The horizontal dotted lines represent the WD inital-final-mass relation for the respective colour-coded progenitor masses (5.0$\mso$ is outside of the y-range of the plot). The blue and green dot represent WD masses that would have had enough time to cool to reach a temperature between 6000K and 8000K.}
        \label{WDcool} 
\end{figure}

The back-of-the-envelope nature of the above calculation implies that the results in Fig. \ref{pd} should be interpreted within an order of magnitude margin. To provide an awareness on the uncertainties associated with this solution curve, we illustrate the effect of an incorrect assessment of the wind velocity, and hence the dynamic crossing time scale at the location of the companion, which was assumed to be 11$\kms$. The shaded zone in Fig. \ref{pd} represents the shift of the solution curve under a 90\% increase and decrease in the assumed wind speed. This is an indicative uncertainty, and by no means does it fully represent the actual error in the curve of solutions. Nevertheless, the generated space of solutions provides an interesting spectrum of companion candidates.

In the low companion temperature regime, and assuming the uncertainties on the black-body radii are substantial, the space of solutions could be consistent with a solar-type star. In the high companion temperature regime only an extremely compact object would be able to explain the observed size of the SiO hole. However, such objects are known to be highly magnetic, and much hotter than the associated temperature on the curve. Hence, such objects can be ruled out. The current analysis indicates that the space of solutions is distinctly consistent with white dwarf (WD) radii \citep{Tremblay2017,Parsons2017,Joyce2018,Romero2019} in the 8000-6000K temperature range. The evolution time of any more massive WD progenitor (few Gyr) to reach the WD state, and the typical WD cooling time (few Gyr) to reach such low temperatures is indeed consistent with typical evolutionary lifetime of ~1$\mso$ star. We used the WD cooling curves given by \citet{Renedo2010} to obtain the cooling time for different WD masses. The results are given in Fig. \ref{WDcool}. The hatched zone represents the cooling time versus WD mass combinations that are consistent with the observations. This result can be used to estimate the mass-range of the tentative WD in order to be able to explain to observations. To this end, we use the combined main-sequence (MS) mass-luminosity relation and MS nuclear time scale relation to estimate the typical MS lifetime of stars with different masses. For a 1.7$\mso$ star such as EP Aqr, this yields a lifetime of 2.65 Gyr, which is the hard upper limit on the WD cooling time (represented by the black vertical line). The actual maximal WD cooling time would then be the lifetime of EP Aqr, minus the lifetime of the WD progenitor. This maximal WD cooling time is represented by the vertical dashed lines: each line representing the WD cooling upper limit for WDs with the respective colour-coded progenitor masses. Finally, combining this with the WD initial-final-mass relation \citep{Cummings2018}, which are plotted as the horizontal dotted lines, we find that the space of solutions is only consistent for WD masses between $\sim$0.65 - 0.8$\mso$, which would have had progenitor masses between 2.1 and 3.3$\mso$.

Hot WDs are typically detected through their strong UV signatures. \citet{Montez2017} show a GALEX spectrum of EP Aqr, which shows no definitive WD signatures inbetween the dominant photospheric emission. This is not surprising considering the strong observational bias associated with observing even much closer, nearly naked WDs, which coincidentally also creates the misconception that WDs are always very hot. Indeed, studies of volume-limited observational samples do confirm that WD temperatures of 5000-6000K are in fact the norm \citep{Hollands2018}, even if they are never detected. Hence, it is uncertain whether any observational campaign in the near future, targetting the detection of EP Aqr's companion will be successful.

For the sake of completion, we note that the space of solutions does not appear to be consistent with the presence of a hot accretion disc around the companion. Such objects have predicted sizes of up to hundreds of solar radii and temperatures up to ten thousand degrees Kelvin \citep{Saladino2018}. However, following the trend in Fig. \ref{pd} an object of such size would have to have temperatures lower than 1000K in order to explain the data. Additionally, such discs are not expected to arise for companions with such large orbital radii (55 AU), since their accretion mechanism would lie well in the Bondi-Hoyle-Lyttleton regime \citep{Hoyle1939,Bondi1944,Bondi1952}, isotropic accretion from wind material within the Hill radius, and would hence be rather inefficient. As discussed by \citet{Homan2018b} a companion mass upper limit of 0.1$\mso$ was derived to explain the non-detection of a main-sequence companion in the continuum. This points the conclusion of our analysis on the nature of the void-inducing companion towards one of two scenarios: Either the extinction along the line-of-sight in the analysis by \citet{Homan2018b} was severely underestimated, in which case the companion could be a brighter main-sequence solar-like star, or the extinction was well estimated, in which case the companion could likely be a white dwarf.

As a final note, we emphasise that the above discussion is based on the distinct assumption that the companion is hot. The possibility of course exists that this assumption is false, and that the companion is of a very different nature. For example, the emission void could also be an extremely cool, dusty and opaque local environment. Such an object would suppresses the radiative excitation of molecules such as SiO locally and, via shadowing of the stellar radiation field, in a tube-like region deep into the radially trailing wind hemisphere. Such an enormous cool, dusty and opaque dust clump has been detected in the inner wind of the red supergiant VY CMa \citep{OGorman2015}, and its origins remain mysterious. For EP Aqr too it remains unclear how such a dust clump would accumulate around the companion. AGB stars are known to occasionally undergo eruptions of a convective cell on their stellar surface, which could generate local conditions that are ideal for the formation of copious amounts of dust. However, it seems unlikely that such a clump of dust would have survived the journey to the location of the companion considering the intricate hydrodynamical streaming lines caused by the wind-companion interaction. The local growth of the clump could have occurred via simple Bondi-Holyle-Lyttleton accretion, but it is not clear what mechanism would have cooled the compressing gas sufficiently to explain the observation.

\subsection{Shape of the polarised light from SPHERE} \label{spherediscuss}

The patterns that emerge in the DoLP maps in the SPHERE data, shown in in Fig. \ref{sphere}, can be explained by three scenarios:

\textbf{Scenario 1:} It is possible that the observed features are generated by directionally confined mass-ejections, caused by the occasional eruptions of the convective cells that dominate the stellar surface. These eruptions are known to generate shocks throughout the loosely bound outer layers of the star that create conditions that are ideal for dust formation \citep{Freytag2017}. The symmetry of the signal in the DoLP maps would hence be a mere coincidence. 

\textbf{Scenario 2:} This same symmetry, however, suggests that the mechanism that determines the dust distribution throughout the inner wind may be less random than the above mentioned scenario suggests. If the dynamical time scale of dust propagation through the inner wind is smaller than the pulsation time scale that fuels the production of dust, and if the dust-producing shocks are randomly distributed in time and space, then the dust may accumulate in a shell surrounding the AGB star. Such features have been invoked in order to explain SED dust excesses \citep{Khouri2015,VanDeSande2018b}. Accounting for the degree of linear polarisation as a function of scattering angle given by the Rayleigh scattering model, such a shell would manifest as a ring in the DoLP maps. This scenario qualitatively explains the data rather well. We note that the DoLP signal becomes dominated by noise beyond $\sim$150 mas, implying that the width of the DoLP signal is by no means a reliable representation for the width of the dust shell. Furthermore, the gaps in the DoLP maps would imply that the shell is incomplete, and that some mechanism has prevented efficient dust formation along the north-west to south-east axis. 

\textbf{Scenario 3:} However, though not perfect, the striking coincidence between the orientation of the symmetry axis of the system (as previously determined by \citet{Homan2018b}) and the patches of decreased DoLP suggests yet a third scenario. Scattering theory predicts that forwards and backwards scattering produces no linearly polarised light, implying that most of the detected scattered light from the pole-on, nearest outflow hemisphere of EP Aqr only contributes a negligible amount to the DoLP signal. The same can be said for the farthest outflow hemisphere. Hence, most of the detected DoLP signal probably originates from the nearly face-on equatorial density enhancement (EDE), in which the dust would form a ring around the AGB star. For a perfect face-on orientation this ring of dust would manifest as a ring in the DoLP maps. However, EP Aqr having a known non-negligible system inclination between 4$^\circ$ and 18$^\circ$ \citep{Homan2018b} the DoLP is expected to decrease along the system's symmetry axis, exactly as observed in the SPHERE data.

\begin{figure}[t]
        \centering
        \includegraphics[width=8.5cm]{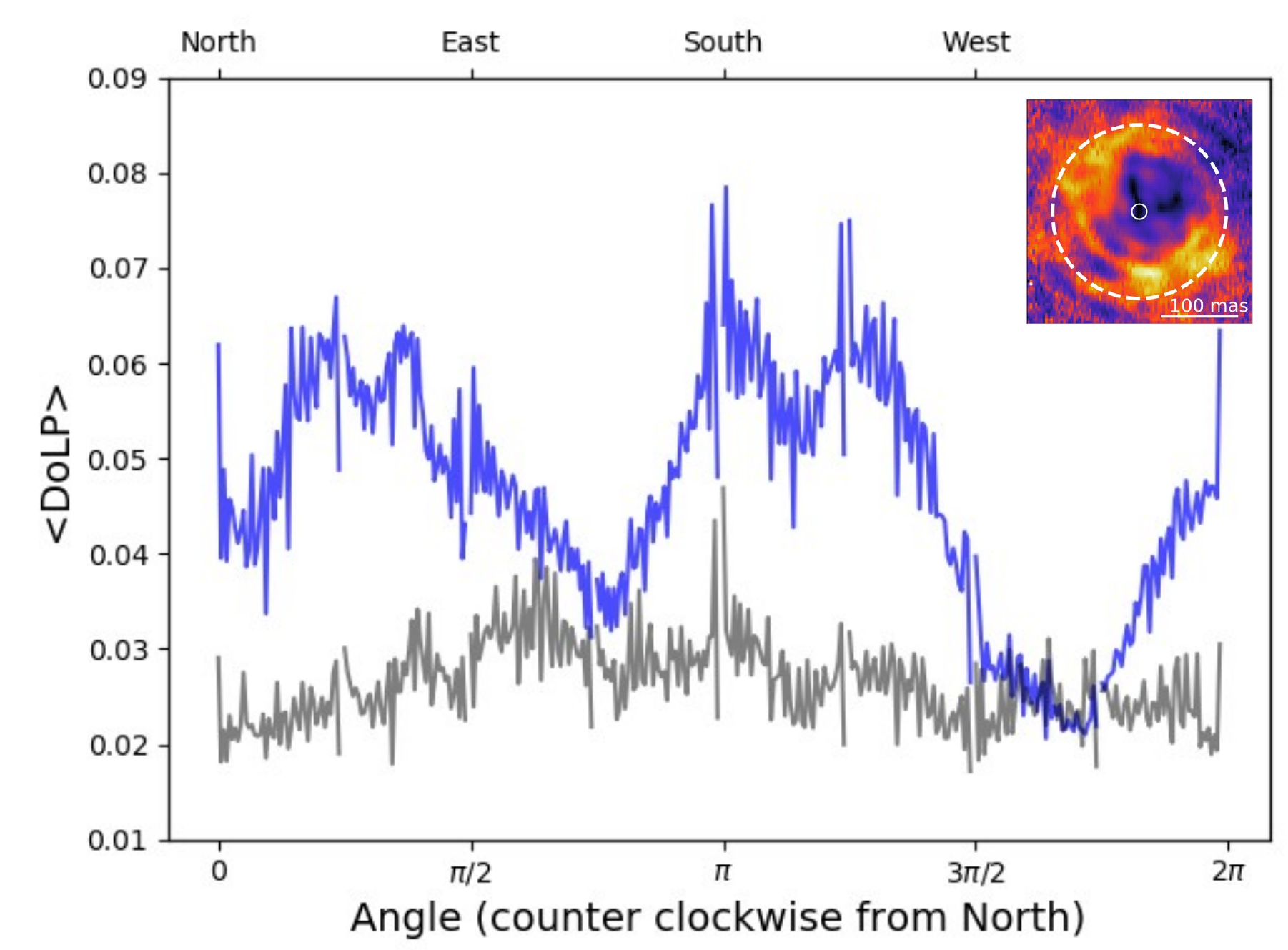}
        \caption{Radially averaged DoLP as a function of angle, shown in blue. The 2$\sigma_{\rm rms}$ noise level is shown in grey. The dashed line in the DoLP image in the top right corner is a visual representation of the outer radius beyond which the DoLP does not contribute to the profile.}
        \label{wiggle}
\end{figure}

Because most physical and geometrical details of EP Aqr's wind are unknown, and because there exist strong degeneracies between some of these properties (such as e.g. dust density and thickness of the dust ring), the current data can by no means be used to provide a quantitative differentiation between the above mentioned scenarios. Nevertheless, considering the morphology of EP Aqr's wind the third scenario provides a satisfying, consistent, and qualitative explanation for the detected shape in the DoLP signal. Furthermore, under the assumption of its validity, it provides additional constraints on the system properties. For one, attributing the decreased DoLP to inclination implies that the EDE morphology reaches all the way down to the inner wind regions that are located approximately 0.1'' away from the stellar surface. The reason for this is not entirely clear if we assume that the tentative companion that is responsible for the SiO void (located at a physical distance of 55 AU) is also responsible for the presence of the EDE. If massive enough, then this companion could in principle bend the otherwise radial streaming lines of the equatorial portion of the wind even more towards the orbital plane, but it is unclear to what degree this can happen for a companion at such a substantial distance, and how important this effect can be near the mass-losing star. Whatever the equatorial wind flattening mechanism may be, the SPHERE data suggest that it may already be operational at distances as close as 0.1'' from the stellar surface. Still under the assumption that the dust that contributes to the observed DoLP maps is located in an EDE surrounding the AGB star, we can use the DoLP signal to provide additional constraints on the symmetry axis of the system and the inclination angle at these length-scales.

In Fig. \ref{wiggle} we show how the DoLP varies as a function of PA in the plane of the sky. The (blue) curve was constructed by averaging the DoLP over a sequence of radial axes originating from the intensity brightness peak position, which we assume to be the centre of the AGB star. To confine the analysis to the region in the field of view which is not dominated by noise (see Fig. \ref{sphererms}) we have set an outer radius determined by the shortest segment (over all radial axes) for which the signal remained above 2 times the noise $\sigma_{\rm rms}$, plotted in grey in Fig. \ref{wiggle}. The blue curve has a periodic aspect: It exhibits two distinct troughs with a $<$DoLP$>$ of around 0.03 and two distinct peaks with a $<$DoLP$>$ around 0.06. Each of these pairs are approximately separated by an angle of $\pi$ radians, they are in geometrically polar opposition. We can estimate the PA of the symmetry axis of the system using the angles where the $<$DoLP$>$ troughs are located, which is at a PA of 138.5$^\circ$. Considering the crudeness of the measurement, this is in remarkable agreement with the result of \citet{Homan2018b} who used the offset between the centres of the blue-shifted and red-shifted wind hemispheres to measure the PA for the symmetry axis of the wind of EP Aqr to be 150$^\circ$, and with the PA of the rotation axis of EP Aqr's the SO$_{\rm 2}$ emission \citep{Homan2018b}, and of the CO emission shown in Fig. \ref{COmom1}.

The peak-to-trough $<$DoLP$>$ ratio provides a crude estimate for the inclination angle of the system. The peaks are quite stable around a $<$DoLP$>$ of 0.06, while the troughs vary quite substantially between 0.035 and 0.024. This translates to a decrease in peak $<$DoLP$>$ between 60\% and 40\%, respectively. Assuming the dust is confined to a cylindrical and flat EDE, then back-of-the-envelope Rayleigh scattering theory predicts scattering angles between $\sim$35 degrees and $\sim$45 degrees. This inclination angle is much higher than the previously determined inclination angle of the system of 4$^\circ$--18$^\circ$ derived from the CO analysis by \citet{Homan2018b}. This difference could point to a fundamental morphological difference between the inner and the outer EDE, but in all probability it can simply be attributed to the crudeness of the inclination estimate. A more reliable estimate for the inclination of the dust disc will be derived from the modelling of the data that will appear in a future work.

\subsection{Tentative tertiary companion} \label{comp2}

The structure observed in the line wings of the CO emission are evidence of an active shaping mechanism which not only affects the global shape of the outflow by forming a spiral-harbouring EDE, but also appears to induce smaller-scale perturbations that manifest as arcs in the outflow hemispheres. In AGB binaries, such arcs are typically attributed to the reflex motion of the AGB star around the centre of mass of the binary system, which produces a spiral shock that due to the geometry of the system can extend to far above and below the orbital plane. For a constant outflow velocity of 11$\kms$, the typical period of the wobbling that would be necessary to generate periodic features at 0.5'' to 0.8'' intervals would be between 30 and 45 years. These could never have been generated by the orbital motions of the AGB binary due to the tentative companion that is thought to be responsible for the SiO emission void.

Alternatively, the presence of an actively accreting companion in close orbit may be able to explain the concentric arcs. \citet{TuanAnh2019} proposed that the augmented speeds observed in the inner wind of EP Aqr (see Fig. \ref{SiOpvd}) could be the manifestation of narrow polar jets that are launched less than $\sim$25 AU away from the central star. They correctly conclude that the tentative companion responsible for the SiO void cannot be the engine that launches these collimated outflows, since it is located too far away from the AGB star. However, a closer companion at a distance of $\sim$10 AU could be a viable candidate to explain these features. Such an object would have an orbital period of $\sim$25 years, which lies quite close to the dynamical crossing time between the observed arcs. This estimate was made under the assumption that the wind speed within 2'' has already reached its terminal value. Any decrease in wind speed would increase the period estimate, bringing it even closer to the dynamical crossing time between two consecutive shells. Collimated streams of gas, ejected nearly perpendicular to the orbital plane, would also form nearly-concentric shell-like features in the wind hemispheres as time progresses due to the mostly radial radiation pressure.

\begin{figure}[t]
        \centering
        \includegraphics[width=8.5cm]{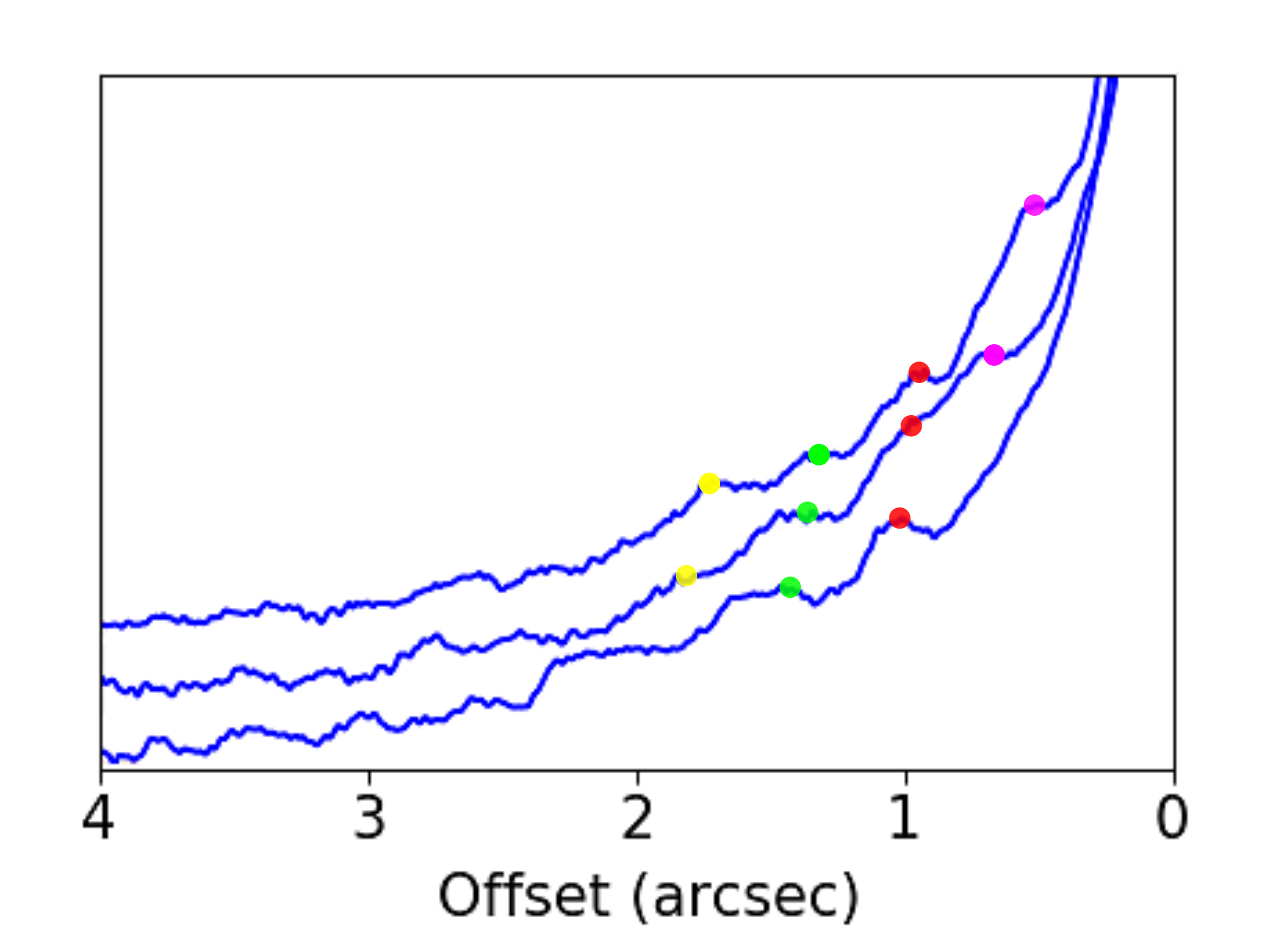}
        \caption{Stacked plot of three different SiO emission profiles taken at different PAs (north to east), vertically shifted by an arbitrary factor to highlight the progression of the bumps. The top profile represents the eastern wing of the profile shown in Fig. \ref{sioprofile}. The middle profile is taken at a PA of 45 degrees, and the bottom profile at a PA of 90 degrees. The dots highlight the bumps to show their outwards shift.}
        \label{progress}
\end{figure}

In addition, the radial SiO emission profile shown in Fig. \ref{sioprofile} exhibits a sequence of 4 consecutive bumps to the east, with a seemingly constant separation of 0.45''. It is by no means clear what the origin of these bumps is. However, dynamical arguments can be used to quickly rule out pulsational shocks or the companion in the void. Hence, we propose that these bumps could also be the signature of spiral shock waves created by the interaction of the wind with a much closer companion. This is supported by the fact that the bumps appear to shift outwards when the PA of the profile is progressively increased, as shown in Fig. \ref{progress}. Assuming the wind has a speed of about 11$\kms$ in the inner wind, the distance between the consecutive bumps would correspond with a period of $\sim$25 years.

Approximately 10\% of solar-type main-sequence stars are found in triple/quadruple systems \citep{Moe2017}. Furthermore, for the currently determined orbital separation of $\sim$10 AU (0.07'') the likelihood for stars with an inital mass greater than 1.5$\mso$ \citep{Fulton2019} (such as EP Aqr \citep{Dumm1998}) of having a >1$M_{\rm Jup}$ planetary companion is higher than 40\%. The presence of such a closer companion could also provide the necessary angular momentum to explain the gas in rotation in the innermost regions of the wind, as was revealed by the moment 1 map of CO shown in Fig. \ref{COmom1}, and a similar map of the SO$_{\rm 2}$ emission shown by \citet{Homan2018b}. And finally, it may also provide a plausible explanation for the reason why the SPHERE data appear to indicate that the EDE of EP Aqr reaches all the way down to 0.1'' distances from the central star, while the most obvious companion candidate resides at 0.4''.

Triple systems are known to be notoriously unstable. However, a multitude of binary systems have been discovered that have planets in stable S-type \footnote{S-type refers to an orbit in satellite configuration. That is, an orbit around one of the two stars in the system. This stands in contrast to P-type orbits, which refer to circumbinary orbits.} orbits \citep{Thebault2015}. It has been shown that a companion star in a wide binary ($>$50 AU) has only a little impact on the formation or distribution of S-type planets, which has been shown to follow the same trend as for single stars \citep{Roell2012}. It is not clear whether such a system remains stable throughout the evolution of the binary, nor whether there exists a set of progenitor orbital configurations that would result in the configuration we observe today, especially considering that EP Aqr has most likely experienced the complete evolution of the secondary from a main-sequence star to the WD phase (see Sect. \ref{wd}).


\section{Summary} \label{summ}

In this manuscript we present the latest high-resolution (0.025'') band 6 ALMA data investigating the wind of the M-type AGB star EP Aquarii, supported by additional SPHERE/ZIMPOL observations of the inner wind though the V, CntH$\alpha$, BH$\alpha$, and Cnt748 filters. These are follow-up observations of previous investigations, which have shown the wind of EP Aquarii to harbour a multitude of morphological complexities: a face-on equatorial density enhancement (EDE) that contains a spiral, a pole-on bi-conical outflow, a highly complex inner wind morphology that resembles the wind-Roche-lobe-overflow models in the literature, a distinct emission void in SiO approximately 0.45'' to the west of the continuum brightness peak (CBP) position, a signal that is indicative of gas in rotation within the inner 0.2'' of the wind, and thin polar high-velocity gas streams. All these features hinted towards the presence of a companion in EP Aqr's wind. The new and better resolved data strongly support this hypothesis: The now resolved SiO emission void has been measured to reside precisely 0.4'' to the west of the CBP position. This position also appears to be at the centre of the complex emission patters that are observed in CO that strongly resemble the typical hydrodynamical gas flow patterns associated with the interaction of a mass-losing AGB star with a companion. In particular, the flow along the L1 and L3 Lagrange points can be identified, before they coalesce into the spiral that manifests at larger length-scales. These flow patterns are consistent with a companion located at 0.4'' to the west of the CBP position. In addition, the central brightness peak of the CO emission also exhibits the signal of rotation discovered in earlier work, but now shown to be confined to the inner 0.1''. Assuming the emission void that emerges in the SiO maps is caused by photodissociation, we tentatively calculated the properties of a black body that could provide sufficient UV photons to explain the size of the void. The space of solutions is particularly consistent with a white dwarf, with a mass between 0.65 and 0.8 $\mso$, though a solar-like companion could not be definitively excluded. The SiO radial profile exhibits four periodic consecutive bumps that cannot be explained by the presence of the above mentioned companion, nor stellar surface pulsations. We tentatively propose that these could be caused by a tertiary companion that that resides as close as 10 AU from the AGB star. The SPHERE data reveal a ring of polarised light around the AGB star, with a radius of $\sim$0.1''. Two symmetrically situated reductions in the degree of linear polarisation signal suggest that the dust responsible for the signal is confined to an inclined plane, with an undetermined thickness, and an inclination of around 45 degrees. These are consistent with the previously determined wind morphology, and show that the EDE in the wind persists to radii below the position of the primary companion candidate, supporting the suggestion that a tertiary companion could already be moulding the wind at smaller radii away from the AGB star.


\begin{acknowledgements}  
We would like to express our sincere thanks to dr. Mark Hollands (Warwick Uni., UK) for his valuable contributions on white dwarf properties. We would also like to thank dr. E. Lagadec and dr. J. Milli for their contributions to the processing and troubleshooting of the SPHERE data. W.H., L.D acknowledge support from the ERC consolidator grant 646758 AEROSOL. E.C. acknowledges funding from the Ku Leuven C1 grant MAESTRO C16/171007. TJM is grateful to the STFC for support via grant number ST/P000321/1. This project has received funding from the European Union's Horizon 2020 research and innovation program under the Marie Sk\l{}odowska-Curie Grant agreement No. 665501 with the research Foundation Flanders (FWO) ([PEGASUS]$^2$ Marie Curie fellowship 12U2717N awarded to M.M.). This paper makes use of the following ALMA data: ADS/JAO.ALMA\#2011.0.00277.S and ADS/JAO.ALMA\#2018.1.00750.S. ALMA is a partnership of ESO (representing its member states), NSF (USA), and NINS (Japan), together with NRC (Canada),  NSC and ASIAA (Taiwan), and KASI (Republic of Korea), in cooperation with the Republic of Chile. The Joint ALMA Observatory is operated by ESO, AUI/NR.A.O, and NAOJ. Based on observations collected at the European Southern Observatory under ESO programme(s) 0102.D-0006(A). This work has made use of the SPHERE Data Centre, jointly operated by OSUG/IPAG (Grenoble), PYTHEAS/LAM/CeSAM (Marseille), OCA/Lagrange (Nice), Observatoire de Paris/LESIA (Paris), and Observatoire de Lyon/CRAL, and supported by a grant from Labex OSUG@2020 (Investissements d’avenir - ANR10 LABX56).
\end{acknowledgements}

\bibliographystyle{aa}
\bibliography{wardhoman_biblio}

\IfFileExists{wardhoman_biblio.bbl}{}
 {\typeout{}
  \typeout{******************************************}
  \typeout{** Please run "bibtex \jobname" to obtain}
  \typeout{** the bibliography and then re-run LaTeX}
  \typeout{** twice to fix the references!}
  \typeout{******************************************}
  \typeout{}
 }

\clearpage
\begin{appendix}

\section{Additional figures}

\begin{figure}[htp]
        \centering
        \includegraphics[width=7cm]{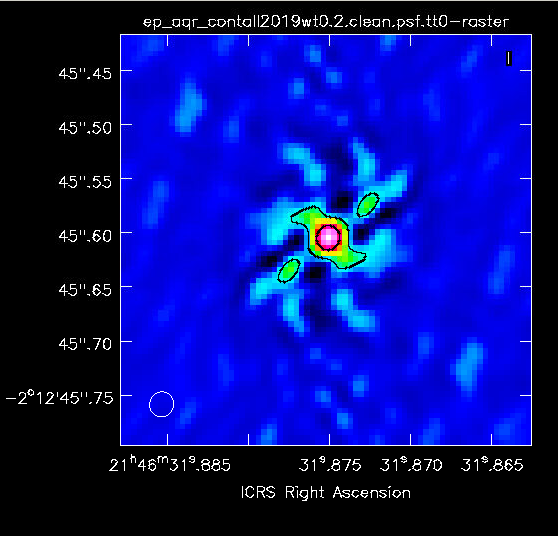}
        \caption{ALMA dirty beam of the combined data.}
        \label{dirty}
\end{figure}

\begin{figure*}[]
        \centering
        \includegraphics[width=18cm]{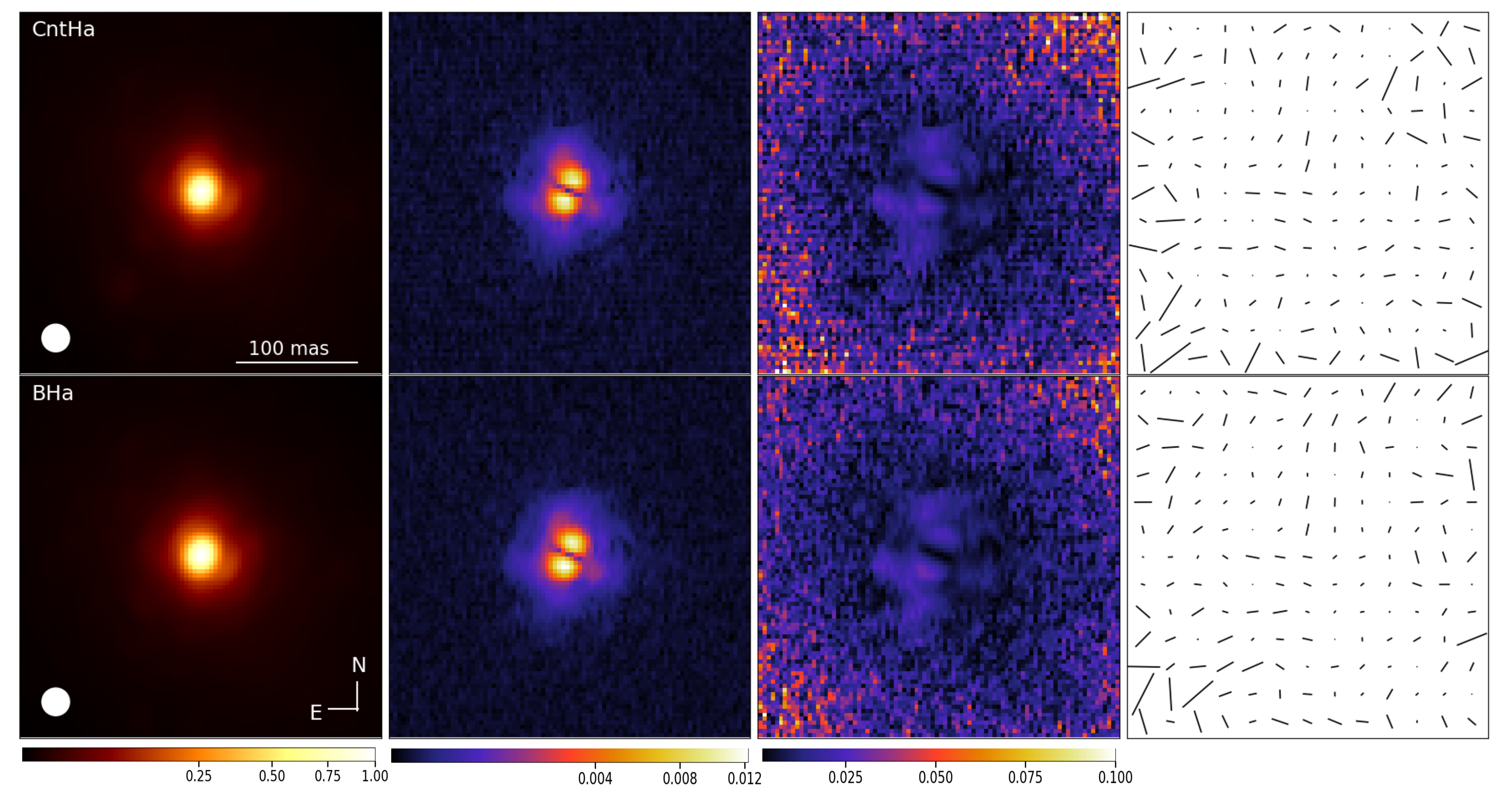}
        \caption{Same as Fig. \ref{sphere} but for the calibrator star HD1921.}
        \label{calib}
\end{figure*}

\begin{figure}[htp]
        \centering
        \includegraphics[width=8.5cm]{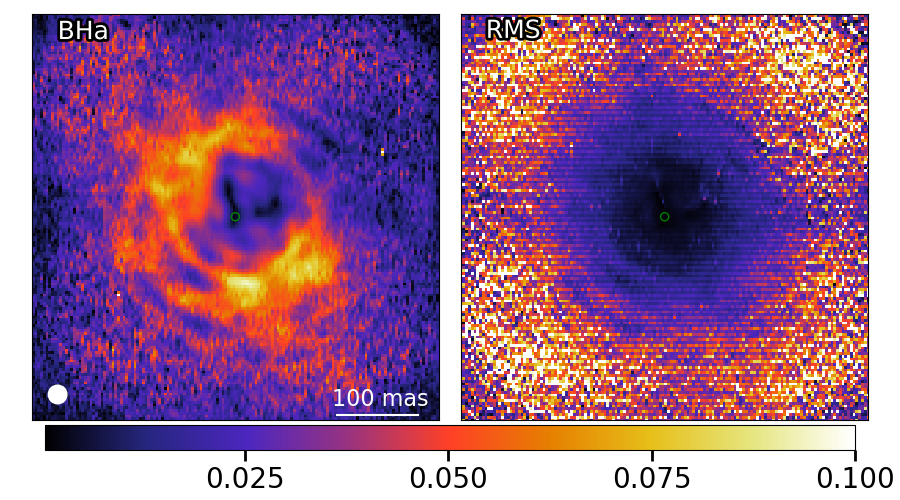}
        \caption{Degree of polarised flux observed through the BHa filter, and associated rms noise map on the right. It is clearly visible that the field of view becomes dominated by noise beyond the features described in Fig. \ref{sphere}.}
        \label{sphererms}
\end{figure}

\begin{figure}[htp]
        \centering
        \includegraphics[width=8.5cm]{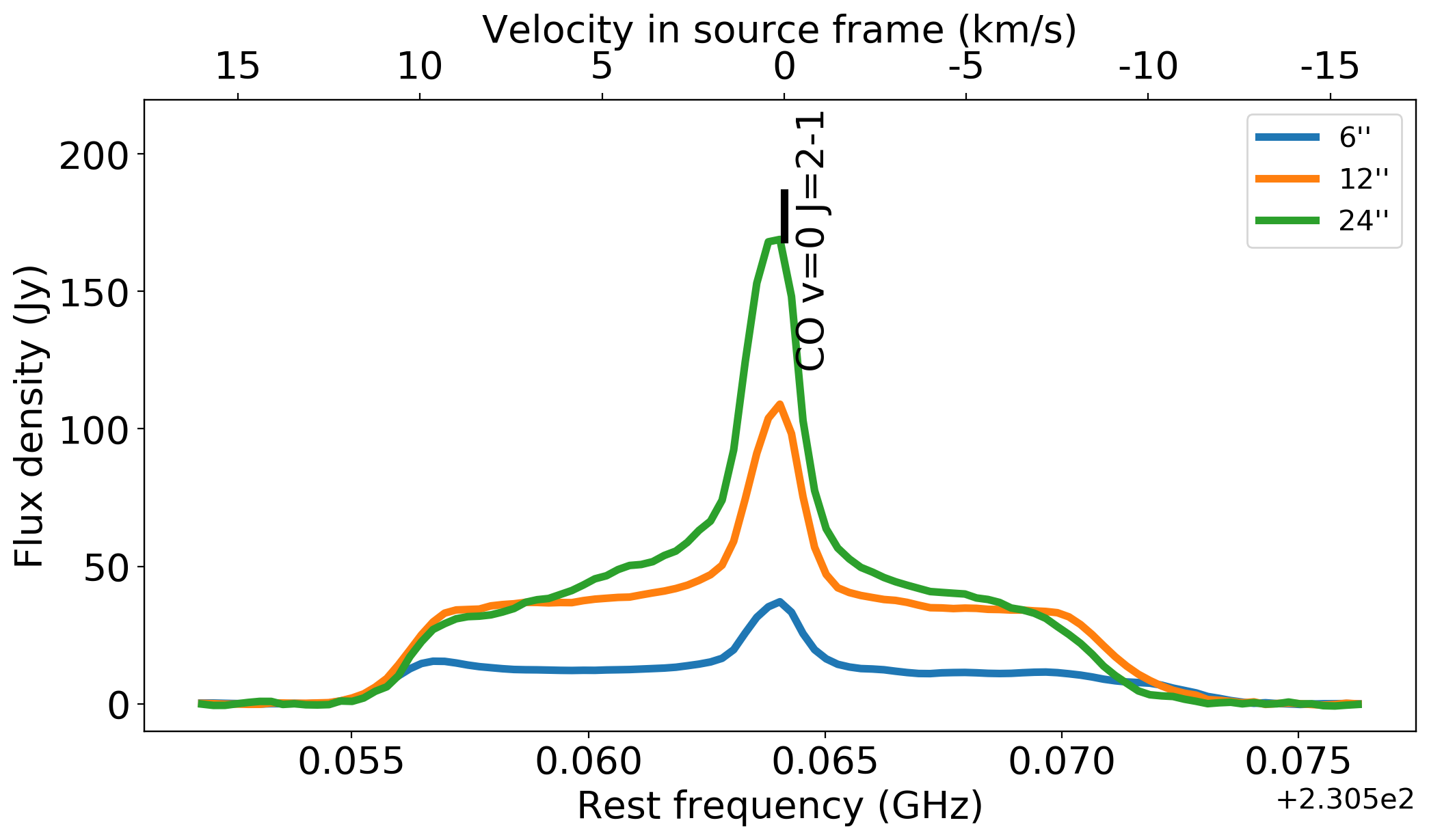}
        \caption{Spectral line of the CO emission for the combined data. Different colours correspond to different circular extraction aperture diameters, as indicated by the plot legend.}
        \label{COline}
\end{figure}

\begin{figure}[htp]
        \centering
        \includegraphics[width=8cm]{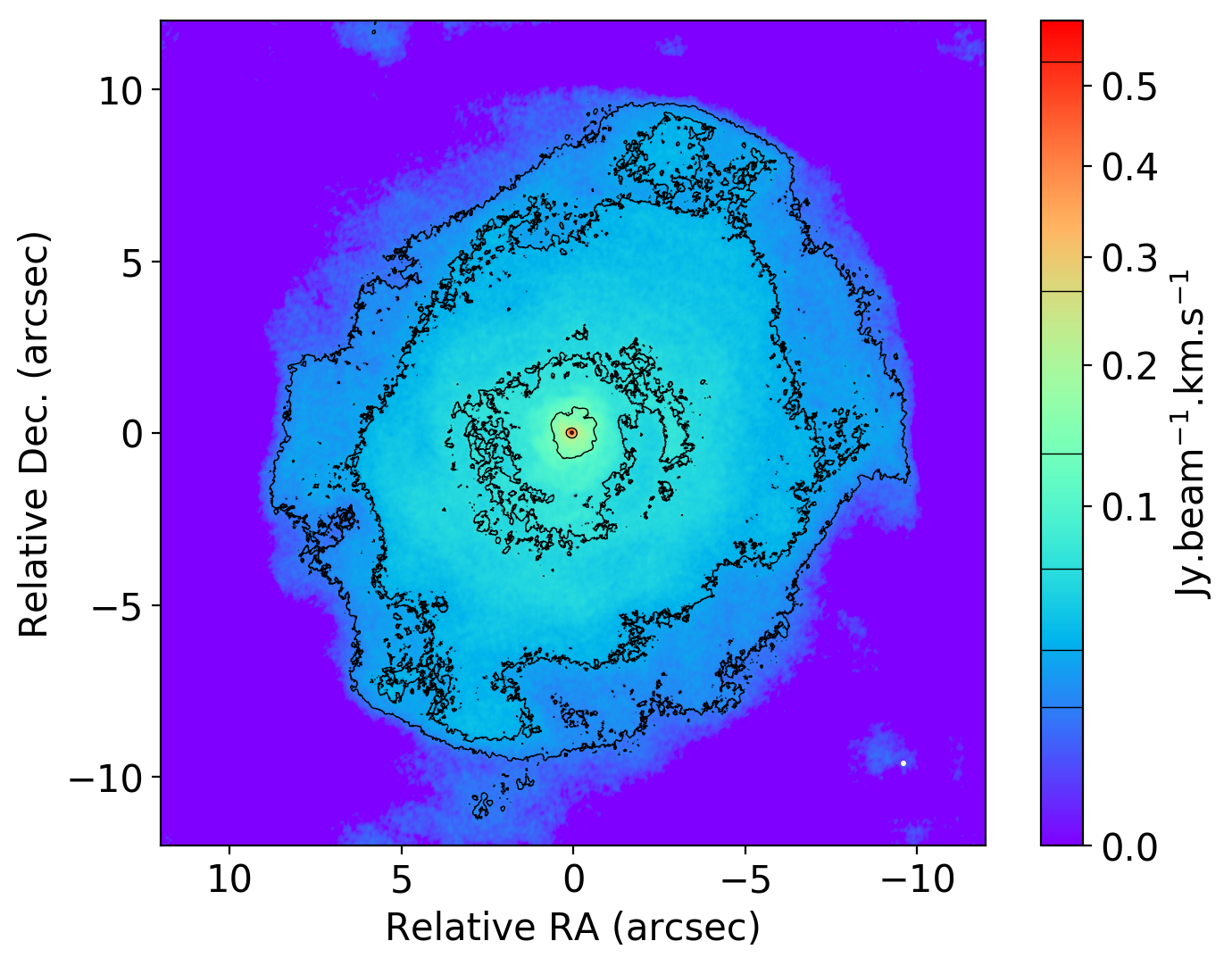}
        \caption{Moment 0 map of the CO emission for the combined data. Contours are plotted et 3, 6, 12, 24, 48, and 96 times the noise rms ($\sigma_{rms}$ = 6.86$\times {\rm 10}^{\rm -4}$ Jy/beam) of the signal-free edge of the map. The ALMA beam size is shown in the bottom right corner.}
        \label{COmom0}
\end{figure}

\begin{figure}[htp]
        \centering
        \includegraphics[width=8cm]{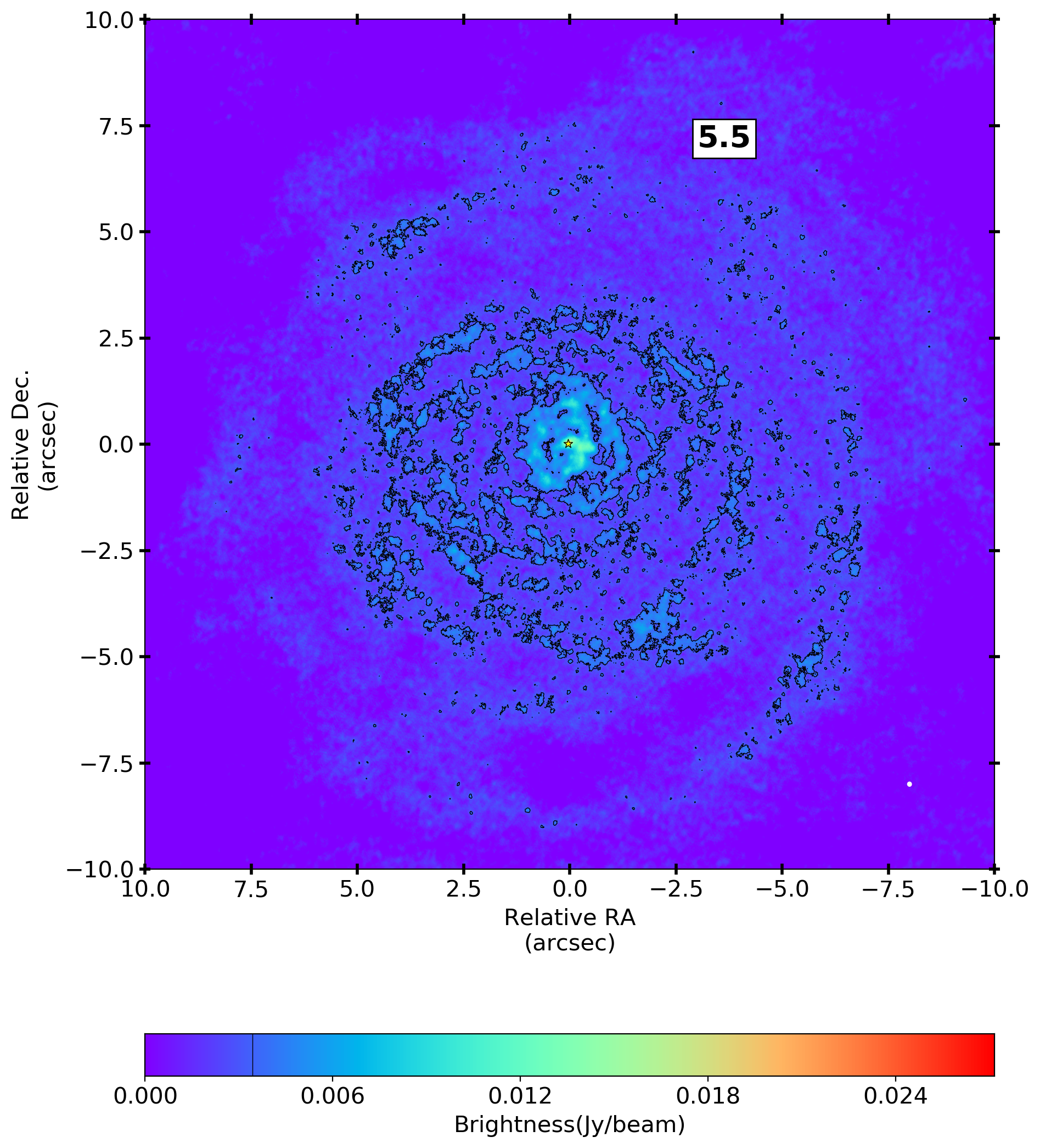}
        \caption{CO channel in the red wing exhibiting the circular arc-like features at the 6 times the rms noise level.}
        \label{COchan84}
\end{figure}

\begin{figure}[htp]
        \centering
        \includegraphics[width=8.5cm]{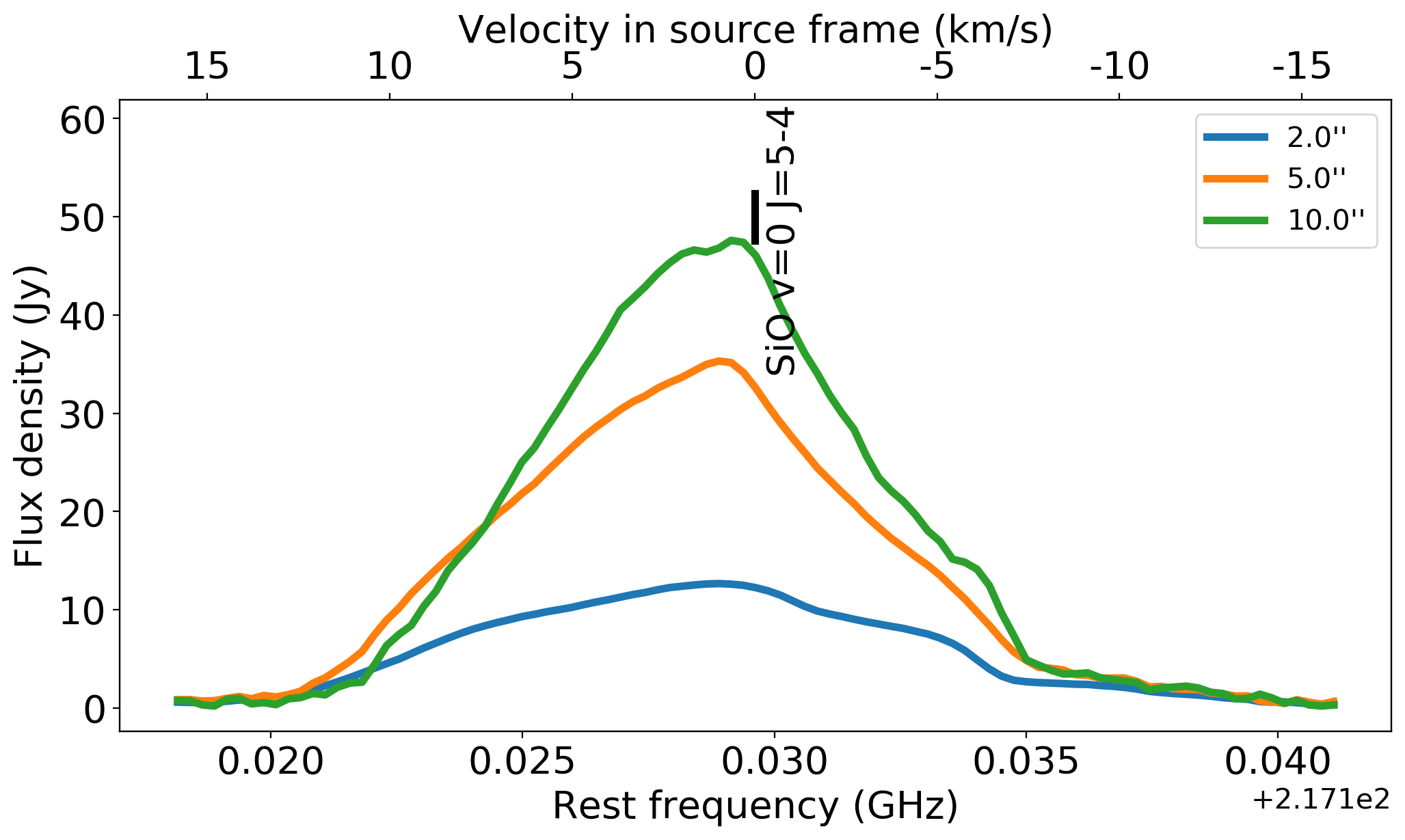}
        \caption{Spectral line of the SiO emission for the combined data. Different colours correspond to different circular extraction aperture diameters, as indicated by the plot legend.}
        \label{SiOline}
\end{figure}

\begin{figure}[htp]
        \includegraphics[width=8.5cm]{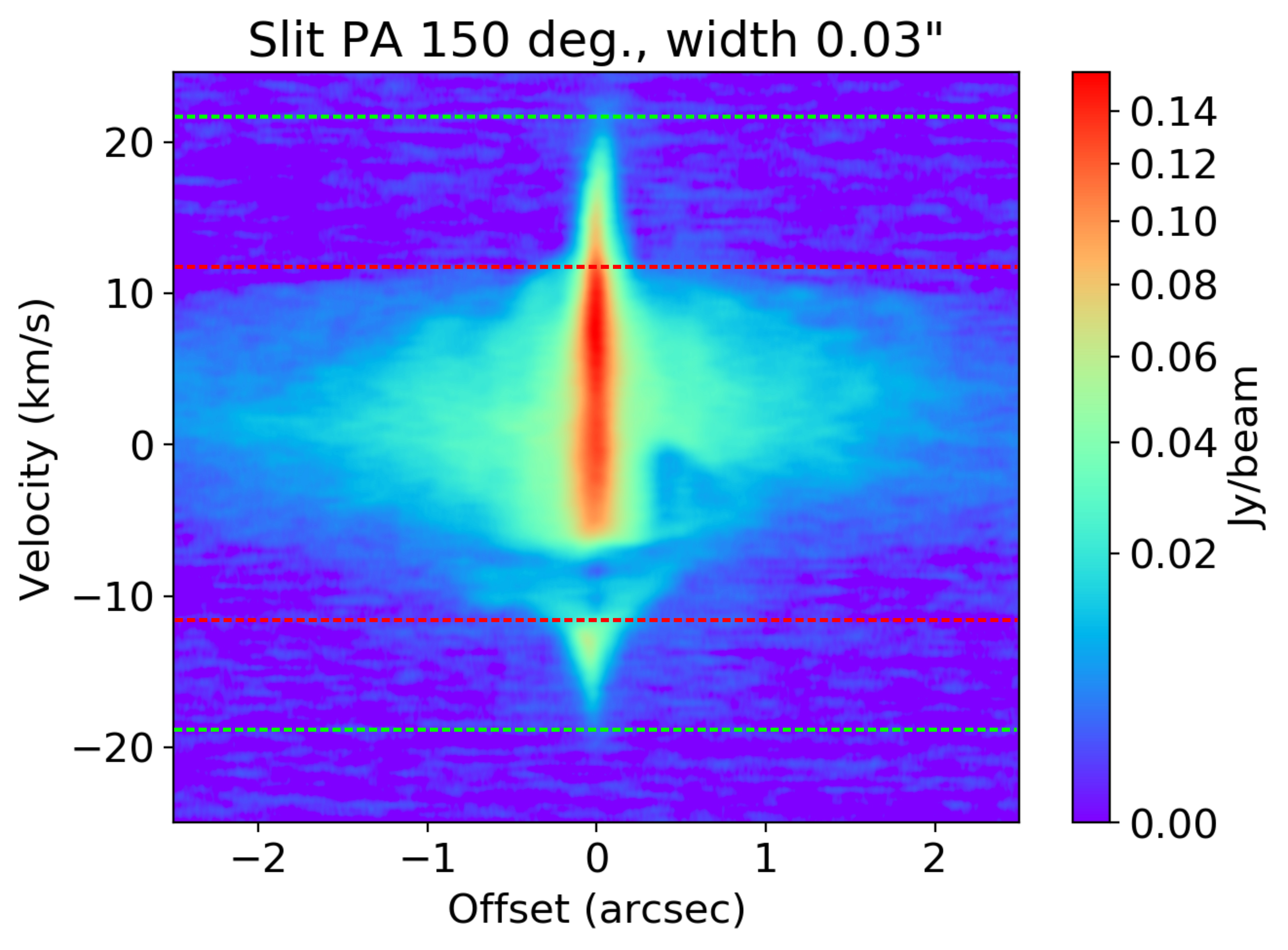}
        \caption{Narrow-slit SiO position-velocity diagram through the CBP position. The red dashed lines are drawn at 12 $\kms$. The green dashed lines show the maximum speed of the compact SiO emission, at -18 and +22 $\kms$}
        \label{SiOpvd}
\end{figure}

\begin{figure}[htp]
        \centering
        \includegraphics[width=8cm]{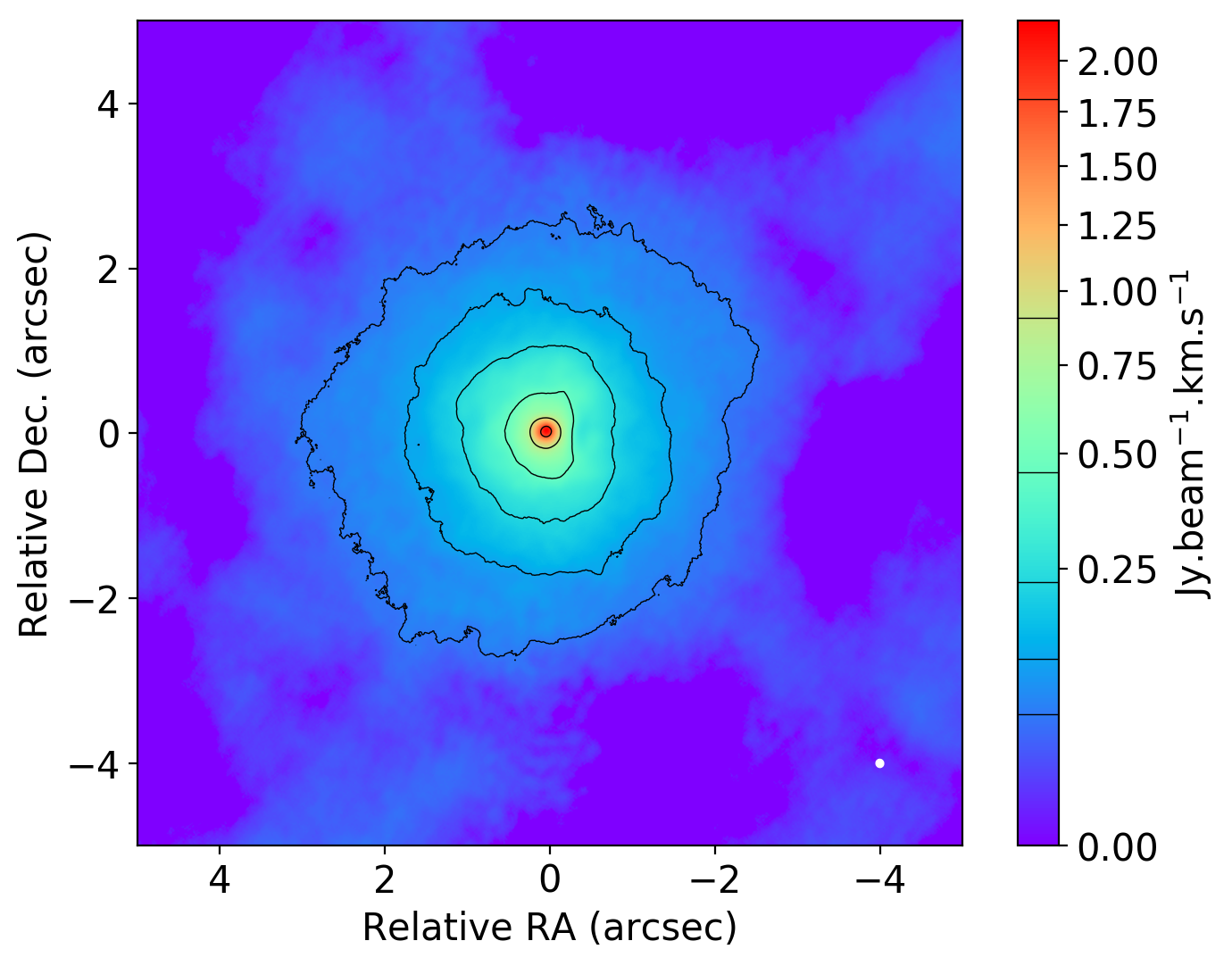}
        \caption{Moment 0 map of the SiO emission for the combined data. Contours are plotted et 3, 6, 12, 24, 48, and 96 times the noise rms ($\sigma_{rms}$ = 1.5$\times {\rm 10}^{\rm -3}$ Jy/beam) of the signal-free edge of the map. The ALMA beam size is shown in the bottom right corner.}
        \label{SiOmom0}
\end{figure}

\begin{center}
\begin{figure*}[htp]
        \includegraphics[width=17cm]{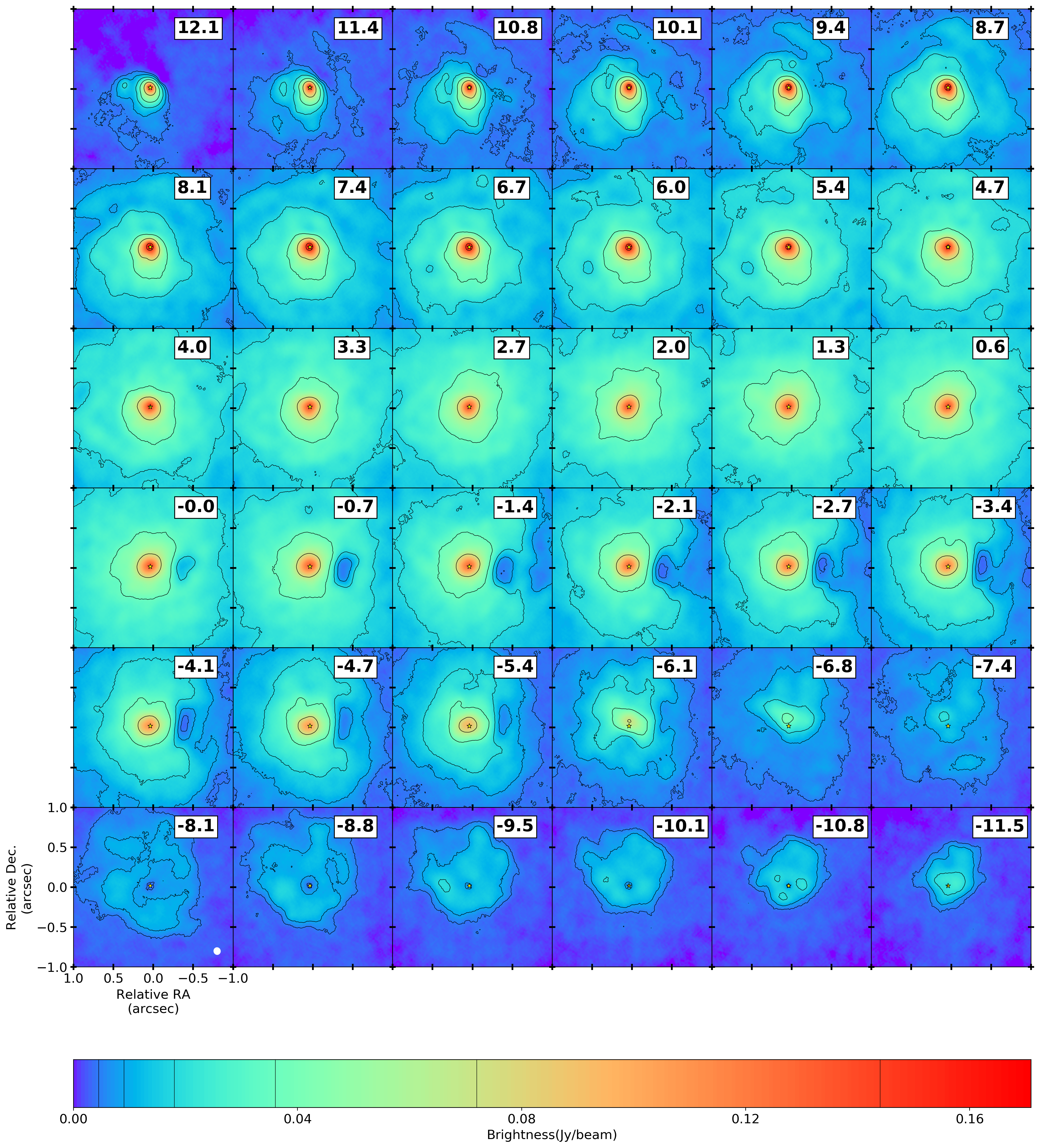}
        \caption{Identical SiO channel maps to Fig. \ref{SiOchan}, but zoomed in on the central 2''$\times$2'' to highlight the resolved void west of the peak brightness position.}
        \label{SiOcz}
\end{figure*}
\end{center}

\begin{figure}[htp]
        \centering
        \includegraphics[width=8cm]{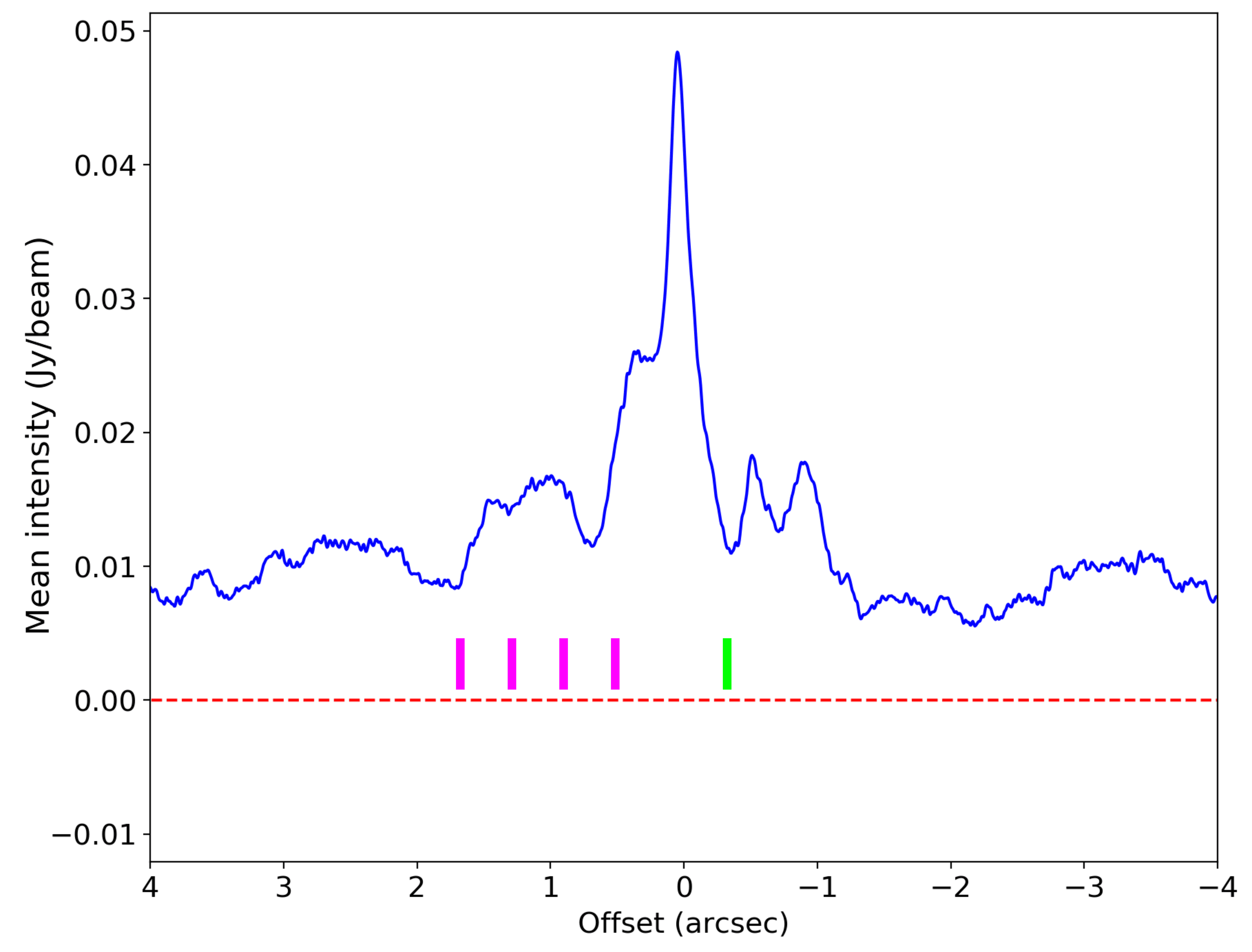}
        \caption{CO emission profile of the central channel, extracted from east to west through the center. The coloured tags indicate the location of the bumps (purple) and the void (green) highlighted in Fig. \ref{sioprofile}.}
        \label{COprofile}
\end{figure}

\begin{figure*}[]
        \centering
        \includegraphics[width=17cm]{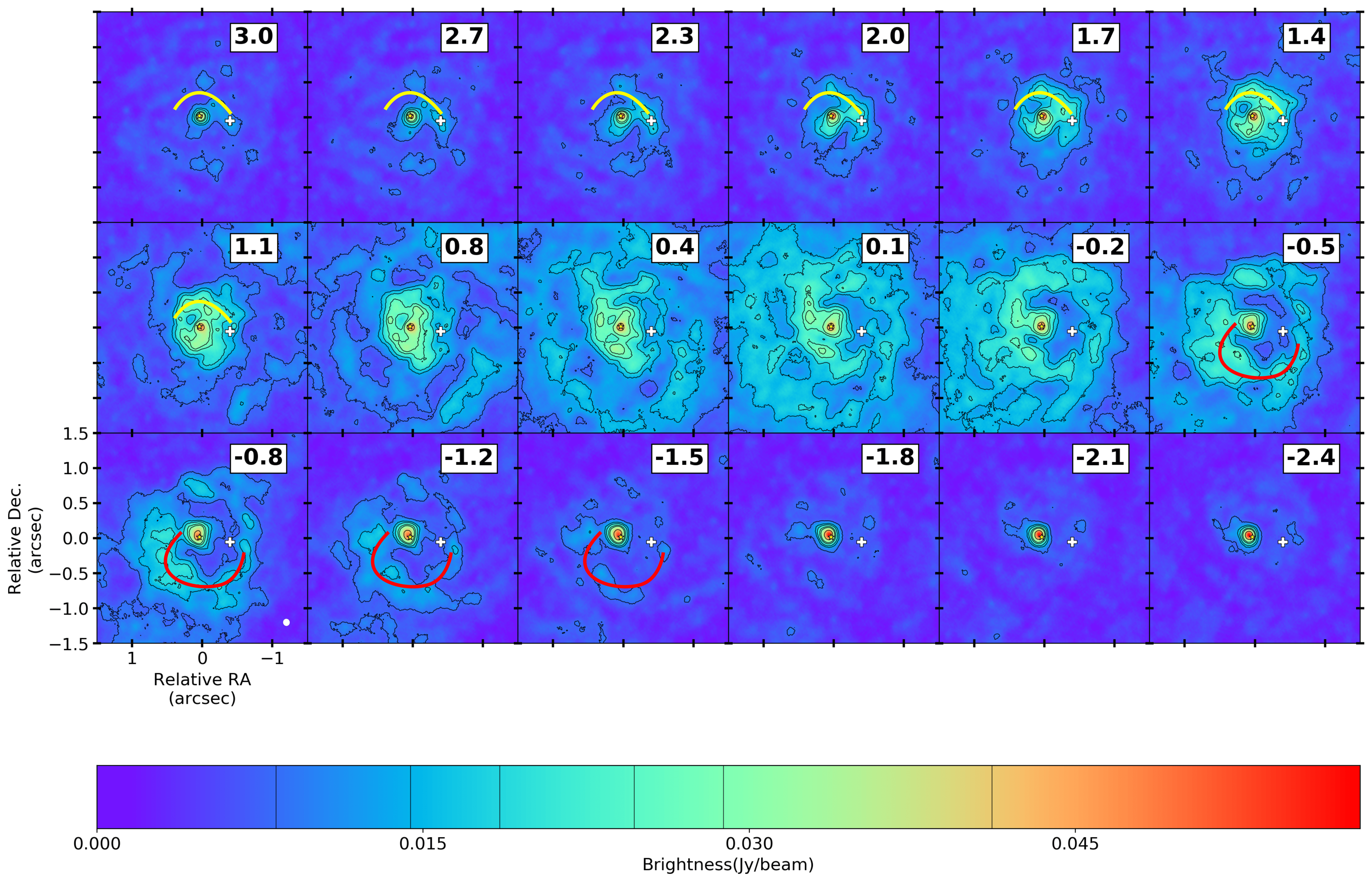}
        \caption{Identical channel maps to Fig. \ref{COcz}, slightly more zoomed in. The white cross on the position of the centre of the SiO emission void to highlight the coincidence between its position and the features in the CO emission. The yellow arc to the north in the red-shifted channels highlights the emission that is associated with the launching of the wake-spiral. The red arc to the south in the blue-shifted challes highlights the emission associated with the reflex-spiral.}
        \label{COhole} 
\end{figure*}

\begin{table}
\caption{Photodissociation rates for SiO assuming a blackbody with a temperature $T_{BB}$, calculated at an arbitrary radius $r_{sc}$ equal to 50 times the radius of the black body.}  
\label{pdtable}   
\centering          
\begin{tabular}{c | c }     
\hline       
\hline      
\noalign{\smallskip}
$T_{BB}$ & Rate \\ 
(K) & (s$^{\rm -1}$) \\ 
\hline 
\noalign{\smallskip}
 5000 & 3.035E-04 \\   
 5500 & 2.031E-03 \\   
 6000 & 9.919E-03 \\   
 6500 & 3.801E-02 \\   
 7000 & 1.204E-01 \\   
 7500 & 3.278E-01 \\   
 8000 & 7.885E-01 \\   
 8500 & 1.714E+00 \\   
 9000 & 3.423E+00 \\   
 9500 & 6.368E+00 \\   
 10000 & 1.115E+01 \\   
 10500 & 1.854E+01 \\   
 11000 & 2.947E+01 \\   
 11500 & 4.506E+01 \\   
 12000 & 6.659E+01 \\   
 12500 & 9.549E+01 \\   
 13000 & 1.333E+02 \\   
 13500 & 1.818E+02 \\   
 14000 & 2.427E+02 \\   
 14500 & 3.178E+02 \\   
 15000 & 4.091E+02 \\   
\hline                 
\end{tabular}
\end{table} 

\end{appendix}
\end{document}